\documentclass[authoryear,3p,11pt]{elsarticle}

\usepackage{amsmath,amssymb,mathtools}

\usepackage{ulem}
\usepackage[usenames,dvipsnames]{xcolor}

\newcommand{\beq}{\begin{equation}}
\newcommand{\eeq}{\end{equation}}

\newcommand{\myeqref}[2]{(\ref{#1}#2)}
\newcommand{\myref}[2]{\ref{#1}(#2)}
\newcommand{\subtag}[1]{\tag{\theparentequation #1}}

\newcommand*\de{\mathop{}\!\mathrm{d}}
\newcommand*{\Oh}{\mathcal{O}}
\newcommand*{\ie}{\emph{i.e.}}
\newcommand*{\eg}{\emph{e.g.}}
\newcommand*{\fvk}{F\"oppl--von K\'arm\'an }

\newcommand*{\Uwork}{\mathcal{U}_{\mathrm{work}}}
\newcommand*{\Ustretch}{\mathcal{U}_{\mathrm{s}}}
\newcommand*{\Ubend}{\mathcal{U}_{\mathrm{b}}}
\newcommand*{\myB}{\mathcal{B}}
\newcommand*{\myF}{\mathcal{F}}
\newcommand*{\myR}{\mathcal{R}}
\newcommand*{\myRs}{\mathcal{R}_s}
\newcommand*{\myT}{\mathcal{E}} 
\newcommand*{\Y}{E_{2D}} 
\newcommand*{\myd}{d}
\newcommand*{\Tpre}{T_{\mathrm{pre}}}
\newcommand*{\Rin}{R_{\mathrm{in}}}
\newcommand*{\Rs}{R_{\mathrm{s}}}
\newcommand*{\Rout}{R_{\mathrm{out}}}

\newcommand*{\rd}{\rho}
\newcommand*{\rr}{\varrho}

\newcommand*{\FT}{F_{\Tpre}}
\newcommand*{\FY}{F_{\Y}}

\journal{Journal of the Mechanics and Physics of Solids }

\begin{document}

\begin{frontmatter}

\title{Indentation of suspended two-dimensional solids: The signatures of geometrical and material nonlinearity}


\author{Thomas G.~J.~Chandler and Dominic Vella}

\address{Mathematical Institute, University of Oxford, Woodstock Rd, Oxford, UK, OX2 6GG.}

\begin{abstract}
The material characterization of ultra-thin solid sheets, including two-dimensional materials like graphene, is often performed through indentation tests  on a flake suspended over a hole in a substrate.  While this `suspended indentation' is a convenient means of measuring properties such as the stretching (two-dimensional) modulus of such materials, experiments on ostensibly similar systems have reported very different material properties. In this paper, we present a modelling study of this indentation process assuming elastic behaviour. In particular, we investigate the possibility that the reported differences may arise from different geometrical parameters and/or non-Hookean deformations, which lead to the system exploring nonlinearities with geometrical or material origins.
\end{abstract}




\end{frontmatter}



\section{Introduction}\label{sec:Intro}

Just as it is natural to test the inflation of a tyre and ripeness of fruit by poking with a finger, a common means of testing the mechanical properties of solids is via indentation tests. While much attention has focussed on the determination of bulk elastic constants via the indentation of a half-space   \cite[see][for example]{Harding1945,Nix1998,Perriot2004,Butt2005,Mckee2011}, a great deal of recent interest has focussed on the use of indentation to determine the mechanical properties of  two-dimensional materials like graphene and molybdenum disulphide \cite[see][for  reviews]{CastellanosGomez2015,Cao2019}. Flakes of such two-dimensional materials are difficult (if not impossible) to manipulate in a tensile testing machine  but may be deposited on a substrate relatively easily. If the substrate on which deposition occurs is patterned with holes, the flake is then `suspended' over the holes: indenting the thin material at a point where it is suspended, for example with an Atomic Force Microscope (AFM),  yields a response that is largely independent of the substrate's mechanical properties (provided that the adhesion between substrate and thin layer is sufficient to guarantee clamping at the hole edge). In particular, indentation yields a force--displacement response that is controlled by the stretching stiffness (or two-dimensional Young's modulus), $\Y$, of the material, together with any residual tension or bending stiffness of the suspended material. 

From the point of view of mechanics, the indentation of suspended  two-dimensional materials has much in common with the indentation of suspended elastic membranes. This problem was studied first by \cite{Schwerin1929}, but has since been extended to account for the effects of   pre-tension \cite[][]{Norouzi2006}, bending stiffness \cite[][]{Wan2003} and indenter geometry \cite[][]{Begley2004, Komaragiri2005}. Nevertheless, these modifications often entail approximate analyses \cite[for example, by assuming an ansatz for the form of the solution, as in][]{Begley2004}, rather than deriving asymptotic results from the full governing equations.

The interest in the mechanical properties of two-dimensional materials has been sparked because of their high material strength, their novel electronic properties, and the interaction between imposed elastic strain and electronic properties \cite[][]{Vozmediano2010,Akinwande2017,Harats2020}; determining the value of the stretching stiffness $\Y$, as well as any pre-existing tension, is therefore the main focus of indentation experiments on suspended two-dimensional materials.

The first   measurement of $\Y$ in graphene was made by \cite{Lee2008}, who reported a value $\Y\approx 340\mathrm{~N/m}$  based on AFM indentation experiments on suspended flakes. While this value is in good agreement with that predicted from first principles \cite[][]{Kudin2001}, values as low as $\Y\approx20\mathrm{~N/m}$ at room temperature \cite[][]{Nicholl2015} and as high as $\Y\approx680\mathrm{~N/m}$ with imposed strain \cite[][]{LopezPolin2017} have been reported. It has also been reported that an optimal number of defects may increase the value of $\Y$ \cite[][]{LopezPolin2015}.

There are many complicating factors involved in the indentation of truly two-dimensional materials like graphene including the importance of thermal crumpling (flexural phonons) and static wrinkles that may both give rise to `hidden area' \cite[][]{Nicholl2017}, the anisotropy induced by the underlying hexagonal lattice \cite[][]{Kumar2015} as well as the possibility of slip at the boundary. In particular, excess membrane area that is hidden in static wrinkles and thermal fluctuations may lead to measurements of the stretching modulus $\Y$ that are small and/or load-dependent simply because out-of-plane deformations are ironed out at low applied stress, as is common when extra material is `buffered-by-buckling' \cite[][]{Vella2019}. This hidden area may have the effect of a strain-dependent stretching modulus and hence be a major cause of the discrepancy in values of $\Y$ reported in the literature, while a stress-dependent stretching modulus of  two-dimensional solids might also arise because of  finite atomic bond lengths. Nevertheless, it has also been pointed out that some inconsistencies may exist in the way that the predictions of classical elasticity theory are used to interpret experimental data \cite[][]{Vella2017,Jia2020}. Here, our aim is to set out clearly the predictions of the standard models of mechanics that are appropriate to thin, isotropic elastic solids as derived asymptotically from the governing equations. We aim  to highlight the potential pitfalls that experimental attempts to characterize material properties by indentation may fall into, focussing in particular on the predictions of these models in the parameter regimes that are of most relevance to two-dimensional solids. A key question in this work will  be how to separate the effect of geometrical nonlinearities (particularly the effect of indenter geometry) from the effect of material nonlinearities (\ie~non-Hookean stress-strain responses). By doing so, we hope that future experiments will be more readily reconciled, and more clearly highlight which experimental results are a consequence of the  unique properties of these unusual solids.

The paper is organized as follows. In \S\ref{sec:modelling} we present our general modelling approach, together with a simple scaling analysis that highlights the variety of possible behaviours; the regimes in which these different behaviours are expected is summarized in Fig.~\ref{fig:regime_cylsph}. We then move on to consider in more detail the effect of indenter geometry with a linear stress-strain relation by considering a cylindrical indenter (\S\ref{sec:cylinder}) and a spherical-tipped indenter (\S\ref{sec:sphere}). In \S\ref{sec:nonlinear_model} we consider the effect of material nonlinearity (\ie~non-Hookean behaviour), before discussing the significance of our results for the experimental determination of elastic constants in \S\ref{sec:discussion} and then summarizing our results and concluding in \S\ref{sec:summary}.

\section{Modelling approach\label{sec:modelling}}

\subsection{Physical model of indentation}\label{sec:Phys_model}

Typical experimental measurements of the mechanical properties of two-dimensional solids by  indentation involve a sheet being suspended over a circular hole, of radius $\Rout$, on an otherwise planar substrate. While the boundary conditions at the edge of the hole are not generally well-controlled, it is usually assumed that the sheet is perfectly clamped at this boundary (\ie~there is no additional radial displacement at the edge as indentation progresses) by the film-substrate adhesion. We shall also  assume that the clamping is perfect.  When the sheet is deposited it is typically subject to a pre-existing  tension, or pre-stress, $\Tpre$, which may be caused by the processing or fabrication of the sheet. We shall assume that the pre-tension is uniform and isotropic and, further, that all deformations are axisymmetric.

A quantity of considerable practical interest is the two-dimensional Young's modulus of the sheet, $\Y$. For a thin, Hookean sheet of thickness $t$, and Young's modulus $E$, $\Y=Et$, but we use $\Y$ throughout so that our theory  describes equally thin elastic sheets and two-dimensional solids. We assume the sheet has a (two-dimensional) Poisson's ratio $\nu$.

\begin{figure}[ht!]
\centering
\includegraphics[width=12cm]{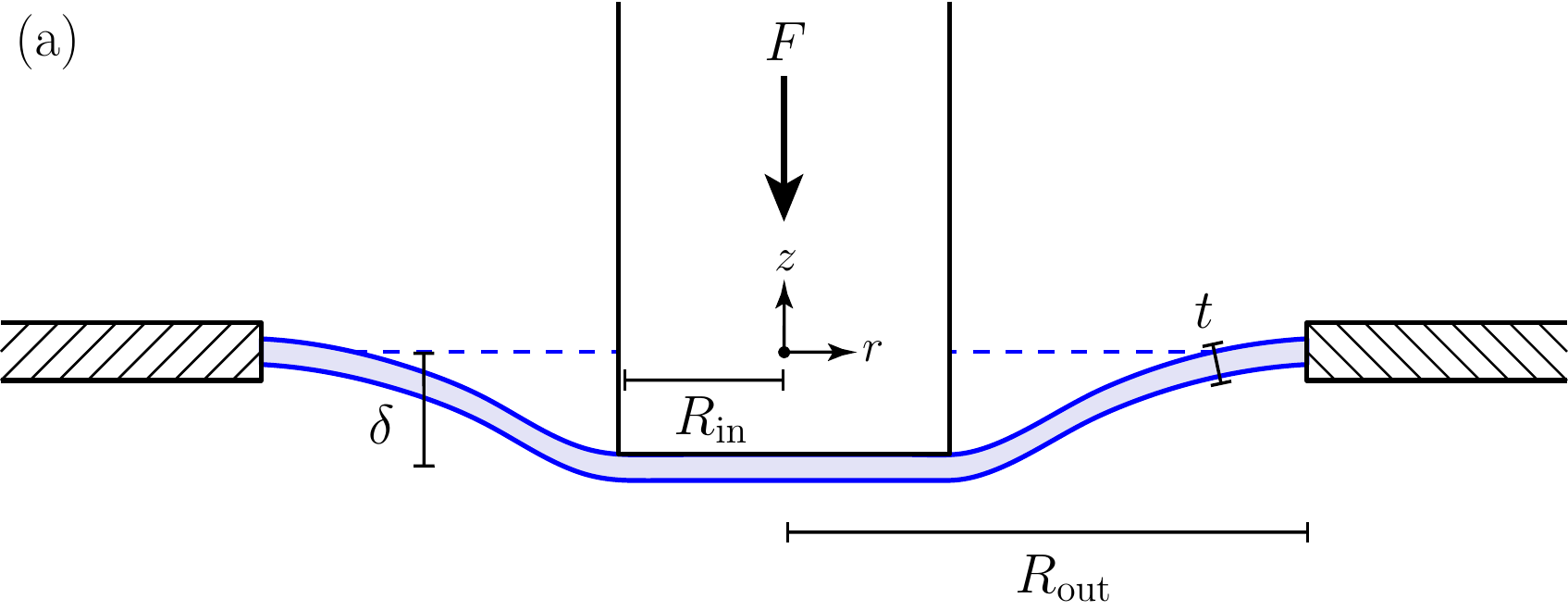}

\vspace*{10pt}

\includegraphics[width=12cm]{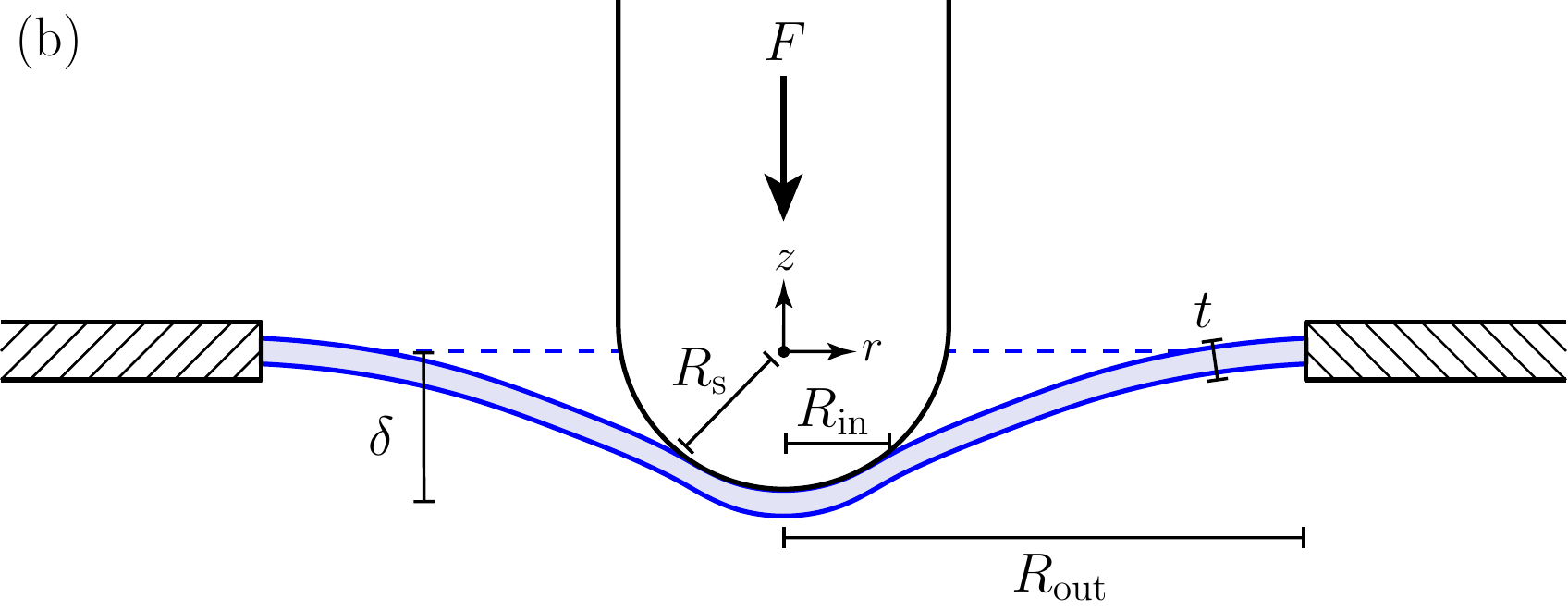}
\caption{Cross-sectional sketch of the indentation of a clamped sheet by: (a)   a cylindrical indenter (or punch) of known radius $\Rin$ and (b) a spherical-capped indenter of known radius of curvature, $\Rs$, but unknown contact radius, $\Rin(\delta)$.
\label{fig:indentation_sketch}}
\end{figure}

 Indentation is typically performed by an Atomic Force Microscope (AFM) tip, which applies the force $F$ required to impose a vertical displacement $\delta$ of the central region (measured relative to the clamped edges). We consider two tip shapes in detail: a cylindrical indenter (of radius $\Rin<\Rout$), which facilitates our analysis, and a spherical tip (of radius of curvature $\Rs<\Rout$), which is more representative of the tips used experimentally \cite[see][for example]{LopezPolin2017}. We shall see that the shape of the indenting tip  can be important for the final interpretation of the force--displacement relationship.  A sketch of the typical experimental setup is shown in Fig.~\ref{fig:indentation_sketch}.

\subsection{Scalings and physical arguments}\label{sec:scalings}

The relationship between the indentation force and depth, $F(\delta)$,  holds important information for understanding  the elastic properties of thin sheets.  To get a first sense of the possible different behaviours of the force--indentation relation, we begin by considering  at a scaling level the various energies in the problem. 

The work done by the indenter, which scales as $\Uwork\sim F\delta$, must be stored predominantly in either the stretching or bending energies of the sheet. To estimate the stretching energy stored within the sheet, we note that the sheet tension has two components: one caused by the pre-tension, $\Tpre$, and another caused by the stretching in response to the imposed strain $\varepsilon\sim (\delta/\Rout)^2$,  which is $\Delta T\sim \Y\cdot\varepsilon$.  (Note that, for small displacements $\delta\ll\Rout$, the strain estimate $\varepsilon\sim (\delta/\Rout)^2$ can be derived from elementary geometrical considerations.) We therefore write $T\sim \Tpre + \Y\varepsilon$ and note that the stretching energy of the sheet must therefore scale according to $\Ustretch\sim\Rout^2(\Tpre+\Y\varepsilon)\varepsilon$. Finally, we note that the bending energy of the sheet scales as $\Ubend\sim B\Rout^2\mathcal{K}^2$, where $\mathcal{K}\sim \delta/\Rout^2$ is the typical sheet curvature and $B$ is the bending modulus (for an isotropic, thin, Hookean solid, $B= \Y t^2/12(1-\nu^2)$, though we use a general bending modulus to account for truly two-dimensional solids, such as graphene, for which the effective value of $B$ may differ significantly from this value\footnote{Typically the bending modulus of graphene is given as around $1.2\mathrm{~eV}$ \cite[see][for example]{Lu2009,Akinwande2017}, or $B\approx 1.9\times10^{-19}\mathrm{~J}$; however, based on the stretching modulus $\Y=340\mathrm{~Nm}$ and thickness $t\approx0.335\mathrm{~nm}$ \citep{Akinwande2017}, the corresponding isotropic solid would have $B=\Y t^2/12(1-\nu^2)\approx3.4\times10^{-18}\mathrm{~J}$. As a result, graphene is 10 times more bendable than would be expected for a corresponding isotropic solid; this reflects the fact that, since it is only a single molecule thick, the usual mechanism for generating a bending stiffness (\ie~differential strain through the sheet thickness) is not relevant for graphene.}).

Before considering the different limits, we note that the work done, $\Uwork$, must equal the sum of these different energies, \ie~$\Uwork\sim \Ubend+\Ustretch$, and hence
\begin{equation}\label{eq:energy_arg}
F\cdot\delta \sim \underbrace{\Rout^2 \Tpre\left(\frac{\delta}{\Rout}\right)^2}_\text{pre-tension} +\underbrace{\Rout^2 \Y\left(\frac{\delta}{\Rout}\right)^4}_\text{stretching} +\underbrace{\Rout^2 B \left(\frac{\delta}{\Rout^2}\right)^2}_\text{bending}.
\end{equation}
Written in this way,  there are three possible dominant balances, depending on which term dominates the RHS of \eqref{eq:energy_arg}: (i)~pre-tension dominated so that $F\sim \Tpre \delta$,  corresponding to linear membrane theory with constant tension \citep[see][for example]{Begley2004, Komaragiri2005, Vella2017}; (ii)~stretching dominated so that $F\sim \Y \delta^3/\Rout^2$,  recovering the scaling of the classical \citeauthor{Schwerin1929} solution \cite[see][]{Schwerin1929,Begley2004, Komaragiri2005, Vella2017}; (iii)~bending stiffness dominated so that $F\sim B\delta/\Rout^2$, and the sheet responds as a classical plate \cite[see][for example]{Timoshenko1959, Wan2003,Komaragiri2005}. 

Taking the pre-tension dominated case as a reference state, natural choices of dimensionless indentation depth, force, and bending stiffness are
\begin{subequations}\label{eq:my_FDB}
\begin{equation}
\myF\coloneqq\frac{F \Y^{1/2}}{2\pi\Tpre^{3/2}\Rout}, \qquad \myd\coloneqq\frac{\delta}{\Rout}\left(\frac{\Y}{\Tpre}\right)^{1/2},
\qquad \myB\coloneqq\frac{B}{\Tpre \Rout^2},\subtag{a--c}
\end{equation}
\end{subequations}
respectively.

Alternatively, one could have taken the bending  or stretching dominated cases as the reference state \cite[][for example]{Komaragiri2005}.  In the applications of current interest (especially for two-dimensional materials) the bending dominated region is of limited interest: for graphene typical values are $B\approx10^{-19}\mathrm{~J}$ \cite[][]{Lu2009} with $\Tpre\approx0.1\mathrm{~N\,m^{-1}}$ and $\Rout\approx1\mathrm{~\mu m}$ \cite[][]{Lee2008} so that $\myB\approx10^{-6}\ll1$. The  pre-tension therefore dominates the bending stiffness and so we  choose a non-dimensionalization that allows the limit $\myB\to0$ to be easily taken.  Similarly, as suggested by \cite{Vella2017}, many recent experiments do not always reach the stretching-dominated (or Schwerin) regime; the non-dimensionalization in \eqref{eq:my_FDB}  allows us to focus on the transition between the dominant balances (i) and (ii). 

\begin{figure}[ht!]
\centering
\includegraphics[width=\textwidth]{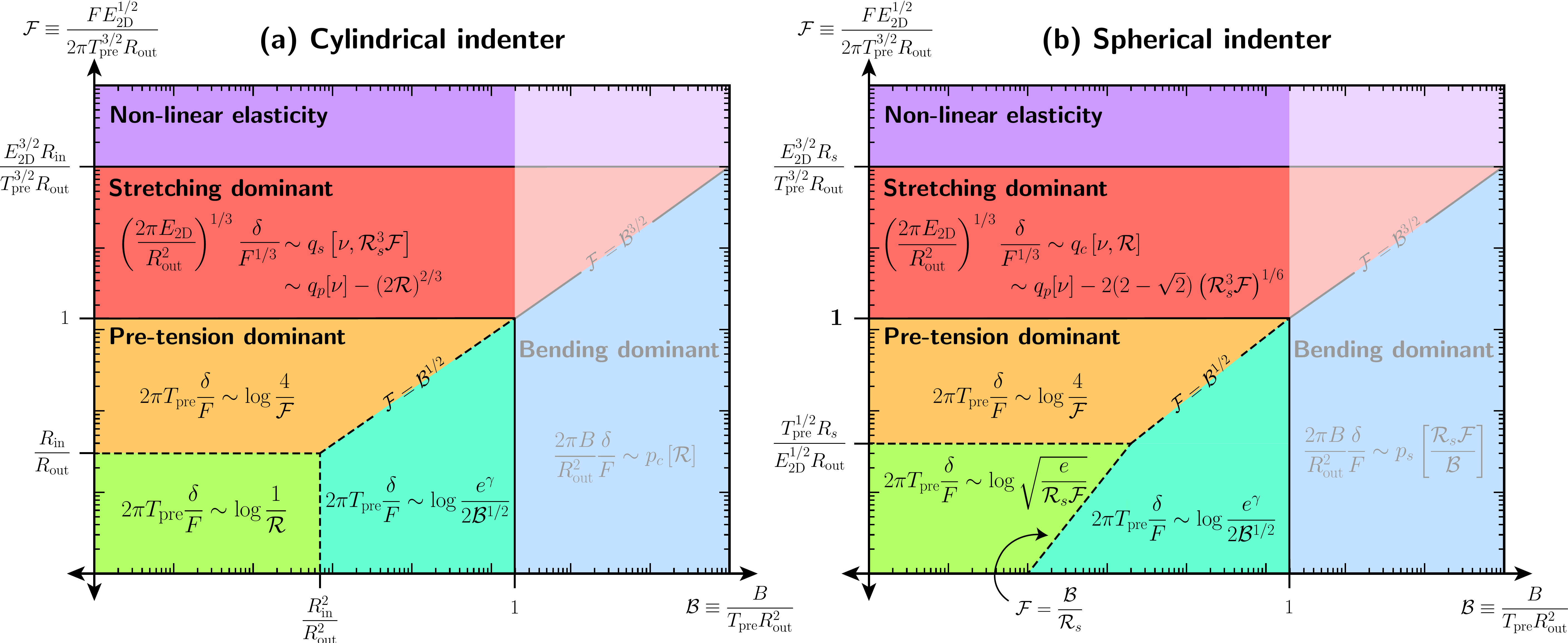}
\caption{Regime diagram for indentation: shown are the regions of the dimensionless-$(\myB,\myF)$-space in which different force--displacement relationships are expected to be observed for (a) a cylindrical indenter of dimensionless radius $\myR=\Rin/\Rout$ and (b) a spherical indenter of dimensionless radius $\myRs=\Tpre^{1/2}\Rs/(\Y^{1/2}\Rout)$. In this paper, we concentrate only on the effects of small bending stiffnesses $\myB\ll 1$. 
\label{fig:regime_cylsph}}
\end{figure}

In dimensionless variables the above dominant balances become: (i)~$\myF\sim\myd$ provided $\{\myF\ll 1,\,\myB\ll 1\}$, (ii)~$\myF\sim \myd^3$  provided $\{\myF\gg \myB^{3/2},\,\myF \gg 1\}$, and (iii)~$\myF\sim  \myB\myd$ provided  $\{\myF \ll\myB^{3/2},\,\myB\gg 1\}$.  These behaviours/regions were the main discussion of \cite{Komaragiri2005} who considered indentation by an idealized point-indenter (with our regions (i)--(iii) respectively corresponding to regions 3--1 in \citealt{Komaragiri2005}). Regions (i)--(iii) are shown in the regime diagrams Fig.~\ref{fig:regime_cylsph} along with subregions (associated with the onset of bending and stretching) and asymptotic results --- which are both established in Sections~\ref{sec:cylinder}~\&~\ref{sec:sphere}. We emphasize that region (i) corresponds to a constant compliance, $d/\myF$, while region (ii) corresponds to a constant `cubic compliance', $d/\myF^{1/3}$. (Discussing compliance, $d/\myF$, rather than stiffness, $\myF/d$, simplifies the analytical results presented here.) We shall be focussed in this paper on understanding the dependence of each of these compliances on the material properties of the system, and the behaviour of the system in-between the asymptotic regimes that correspond to regions (i) and (ii).

From these simple energy arguments, one might assume that the indenter's geometry has little effect on the response $\myd(\myF)$. This assumption has been made implicitly across a range of experimental work  \cite[][to name a few]{Lee2008,CastellanosGomez2015,LopezPolin2015, LopezPolin2017} --- in a complex experimental setup, applying the point-indenter `solutions' to experimental data allows progress to be made. In this paper, we investigate the circumstances in which the radius of contact, $\Rin$, and the shape of the indenter matter.

For a cylindrical indenter we therefore introduce the dimensionless radius 
\begin{equation}\label{eq:my_rad:cyl}
\myR\coloneqq\frac{\Rin}{\Rout}. 
\end{equation} 
For a sphere, it is not immediately clear what  length scale should be used to measure the sphere size; here we use the tensile length $\Rout(\Y/\Tpre)^{1/2}$, which is used to rescale vertical deflections; we thus introduce the dimensionless sphere radius:
\begin{equation}
\label{eq:my_rad:sph}
 \myRs\coloneqq\frac{\Rs }{\Rout}\left(\frac{\Tpre}{\Y}\right)^{1/2}. 
\end{equation}

\subsection{Moderate strains and rotations}\label{sec:moderate_strains}

The energy arguments described in the last section  assumed that the stress and strain were linearly related, \ie~that the material remains Hookean throughout.  This assumption even fails before indentation occurs if the pre-strain $\sim\Tpre/\Y=O(1)$ or during indentation if the stretching induced strain $\sim F/(r \Y)=O(1)$ for $\Rin\leq r \leq \Rout$. In dimensionless variables these two conditions are equivalent to requiring:
\begin{subequations}\label{eq:breakdown_fvk}
\begin{equation}
\myT\ll 1 \quad \text{and} \quad \myF \ll \frac{\myR}{\myT^{3/2}},
\subtag{a,b}
\end{equation}
\end{subequations}
for a Hookean response, where we have introduced the new dimensionless variable
\begin{equation}\label{eq:myT}
\myT \coloneqq \frac{\Tpre}{\Y}, 
\end{equation}
as a measure of the pre-strain in the sheet caused by the pre-tension $\Tpre$. (The variable $\myT$ is only required for the nonlinear elastic model presented in \S\ref{sec:nonlinear_model}.) In the case of a spherical indenter (for which the radius of contact $\myR$ is unknown) an analogous bound to \myeqref{eq:breakdown_fvk}{b}, can be formulated by noting that our detailed analysis in \S\ref{sec:sphere} shows that $\myR \sim(\myRs^3\myF)^{1/4}$, eq.~\eqref{eq:cubic_sph_R}, so that the linear analysis holds provided that $\myF \ll \myRs/\myT^{2}$.

When the strains remain small, it is possible to make reasonable amounts of analytical progress \cite[see][for example]{Vella2017}. However, much of the recent interest in graphene has focussed on whether its material properties change measurably with strain (see \eg~\citealt{Nicholl2015,LopezPolin2017}). We shall, therefore, be interested here in presenting models of indentation in which the material behaviour becomes nonlinear at some point during indentation. In \S\ref{sec:nonlinear_model} we consider the effects of large strains/slopes and nonlinear constitutive relationships using the theory of finite elasticity. For now, however, we focus on describing the linearly elastic behaviour more fully.

\section{Cylindrical indentation of a linearly elastic sheet}\label{sec:cylinder}	

We begin by considering the case of a cylindrical indenter, for which the specification of the problem (especially the boundary conditions) is relatively simple.

\subsection{F\"oppl--von K\'arm\'an formulation}\label{sec:cylinder_fvk}
We initially confine our attention to linear elasticity with the Kirchhoff assumptions (\ie~small strains and plate rotations), so that the F\"oppl--von K\'arm\'an (FvK) equations hold  \citep{Mansfield2005}.  The axisymmetric FvK equations link the out-of-plane displacement of the sheet, $z(r)$, to the stress via a stress  potential  $\psi(r)$, which is defined such that the principal stresses are $\sigma_{rr}=\psi(r)/r$ and $\sigma_{\theta\theta}=\psi'(r)$, thereby ensuring that the in-plane equation holds automatically \cite[][]{Mansfield2005}.  The out-of-plane force balance and the compatibility of strains condition may both be integrated once to give:
\begin{subequations}\label{eq:fvk}\begin{align}
Br \frac{\de \,}{\de r}\left[\frac{1}{r}\frac{\de\,}{\de r} \left(r \frac{\de z}{\de r}\right)\right]&= \psi \frac{\de z}{\de r} - \frac{F}{2\pi},\label{eq:fvk_1}\\
r\frac{\de\,}{\de r}\left[\frac{1}{r} \frac{\de \,}{\de r}\left(r\psi\right)\right] &=  -\frac{\Y}{2}\left(\frac{\de z}{\de r}\right)^2,\label{eq:fvk_2}
\end{align}\end{subequations}
respectively. Note that, since we are assuming linear elasticity, the deformed and undeformed configurations are interchangeable; the radial coordinate in the sheet is thus denoted $r\in[\Rin,\Rout]$. 

At the outer-rim, $r=\Rout$, we assume a perfectly clamped boundary:
\begin{subequations}\label{eq:bc_outer}\begin{equation}
z(\Rout) =0, \qquad z'(\Rout)=0,  \qquad \psi'(\Rout)- \nu \frac{\psi(\Rout)}{\Rout} = (1-\nu)\Tpre. \subtag{a--c}
\end{equation}\end{subequations}
Here \myeqref{eq:bc_outer}{a,b} are geometric conditions of zero vertical displacement and slope, while \myeqref{eq:bc_outer}{c} ensures  the radial displacement is fixed to be that caused by the initial isotropic tension $\Tpre$.

At the inner-rim, $r=\Rin$, we assume a perfect-slip boundary (in reality there may be a small amount of adhesion/friction between the indenter and sheet):
\begin{subequations}\label{eq:bc_cyl}\begin{equation}
z(\Rin) = -\delta, \qquad z'(\Rin) = 0, \qquad \psi'(\Rin)  - \frac{\psi(\Rin)}{\Rin} =0. \subtag{a--c}
\end{equation}\end{subequations}
Here \myeqref{eq:bc_cyl}{a,b} are geometric conditions of continuous vertical displacement and slope; note that \myeqref{eq:bc_cyl}{c}  comes from a force balance with the inner (known-geometry) solution [$\psi(r)\propto r$], explaining why \myeqref{eq:bc_cyl}{c} is independent of Poisson ratio $\nu$ in contrast to \myeqref{eq:bc_outer}{c}. Throughout this paper we shall assume that the sheet and indenter remain in contact for $r\leq\Rin$.

\subsubsection{Non-dimensionalization}

We use the dimensionless variables  suggested by the discussion of energy and scalings in \S\ref{sec:scalings}; in particular, we define
\begin{subequations}\label{eq:dimless_var}
\begin{equation}
\rd \coloneqq \frac{r}{\Rout}, \qquad \Psi \coloneqq\frac{\psi}{\Tpre\Rout}, \qquad Z \coloneqq\frac{z}{\Rout} \left(\frac{\Y}{\Tpre}\right)^{1/2}.\subtag{a--c}
\end{equation}
\end{subequations}
Substitution of \eqref{eq:dimless_var} into \eqref{eq:fvk}--\eqref{eq:bc_cyl} gives the dimensionless system
\begin{subequations}\label{eq:fvk_dless}\begin{align}
\myB \rd \frac{\de \,}{\de \rd}\left[\frac{1}{\rd}\frac{\de\,}{\de \rd} \left(\rd \frac{\de Z}{\de \rd}\right)\right] &=  \Psi \frac{\de Z}{\de \rd}  -\myF,\label{eq:fvk_dless1}\\
\rd\frac{\de\,}{\de \rd}\left[\frac{1}{\rd} \frac{\de \,}{\de \rd}\left(\rd\Psi\right)\right] &= - \frac{1}{2}\left(\frac{\de Z}{\de \rd}\right)^2,
\end{align}\end{subequations}
for $\myR\leq \rd \leq 1$, subject to the boundary conditions
\begin{subequations}\label{eq:bc_cyl_dless}\begin{align}
 Z(1) &= 0, & Z'(1)& = 0, & \Psi'(1)- \nu \Psi(1) &= 1-\nu,  \subtag{a--c}\\
  Z(\myR) &= -\myd, &  Z'(\myR) &= 0, & \Psi'(\myR)- \frac{\Psi(\myR)}{\myR}&= 0.\subtag{d--f}
\end{align}\end{subequations} 
(Recall that $\myF$, $\myB$, and $\myR$ are defined in equations \myeqref{eq:my_FDB}{a}, \myeqref{eq:my_FDB}{c}, and \eqref{eq:my_rad:cyl}, respectively.)

For given dimensionless parameters $\nu$, $\myB$, $\myF$, and $\myR$, \eqref{eq:fvk_dless} subject to \eqref{eq:bc_cyl_dless} may be solved by a standard numerical integrator (in our work we used \texttt{bvp4c} in \texttt{\textsc{Matlab}}). To make analytical progress we consider separately two asymptotic limits that allow for simplifications: (i)~small deflections from the pre-stretched state and (ii) negligible bending stiffness. In \S\ref{sec:results_B>0_cyl} we consider the first of these, by considering perturbations to the isotropic pre-tensed state ($\psi\sim \Tpre r$, $\Psi\sim \rho$),  thereby extending the work of \eg~\cite{Jennings1995,Wan2003}, while in \S\ref{sec:results_B=0_cyl} we consider the second case by considering the zero bending stiffness limit ($\myB=0$), similarly to \eg~\cite{Bhatia1968,Vella2017}. This analysis allows us to reproduce some previously known results in a systematic way, whilst also uncovering new results in some regimes; we discuss the broader context of these results as they are derived.

\subsection{Small indentation forces $(\myF\ll\max\{\myB^{1/2},\myR\})$}\label{sec:results_B>0_cyl}

For sufficiently small indentations, the stretching of the sheet is negligible compared to the isotropic pre-tension and the (also small) bending stiffness, $\myB\ll 1$. To investigate how bending and pre-tension interact, we follow the approach of \eg~\cite{Vella2012,Box2017}, and linearize  the governing equations \eqref{eq:fvk_dless} about the initial pre-tensed state: we let $\Psi(\rho)\sim\rho+\tilde{\Psi}(\rho)$ and $Z(\rho)\sim0+\tilde{Z}(\rho)$ for $\tilde{\Psi},\tilde{Z}\ll 1$. The two equations \eqref{eq:fvk_dless} thus reduce to a single third-order differential equation for $\tilde{Z}$ that can be solved by a linear combination of logarithms and modified Bessel functions (see \ref{app:small_forces}). Applying the boundary conditions \eqref{eq:bc_cyl_dless} we obtain an explicit  relationship between the force and indentation depth that is linear (\ie~$\myF\propto \myd$). We express this relationship  through the (constant) compliance $\myd/\myF$:
\begin{equation}\label{eq:Bessel_cylsol}
\frac{\myd}{\myF} =  \log\frac{1}{\myR}+\frac{\hat{K}_1\hat{I}^R_0+ \hat{I}_1\hat{K}^R_0 +\myR\left(\hat{K}_0\hat{I}^R_1 +\hat{I}_0\hat{K}^R_1\right)-2\myB^{1/2}}{\hat{I}^R_1 \hat{K}_1 - \hat{I}_1\hat{K}^R_1}\frac{\myB^{1/2}}{\myR},
\end{equation} 
where
\begin{subequations}\label{eq:myIK}
\begin{align}
\hat{I}_j \coloneqq I_j\left(\myB^{-1/2}\right), &\qquad \hat{I}^R_j \coloneqq I_j\left(\myR\myB^{-1/2}\right), \subtag{a,b}\\
\hat{K}_j \coloneqq K_j\left(\myB^{-1/2}\right), &\qquad \hat{K}^R_j \coloneqq K_j\left(\myR\myB^{-1/2}\right), \subtag{c,d}
\end{align}
\end{subequations}
and $I_j(x)$ and $K_j(x)$ are the $j$th-order modified Bessel functions of the first and second kind, respectively \cite[][]{Abramowitz1964}.

Equation \eqref{eq:Bessel_cylsol} corresponds to a constant compliance regime: the compliance $\myd/\myF$ is a function of the indenter radius, $\myR$, and bending stiffness, $\myB$, only. One could use this solution as an explicit formula to describe the  small indentation compliance of a tense clamped plate. As a tool for  inferring the bending stiffness from an experimental measure of the compliance, however, the complexity of this equation is daunting since the various dimensionless quantities are coupled within the Bessel functions. Instead, recall that we are specifically interested in the limit of small bending stiffnesses $\myB\ll 1$ and note, from \myeqref{eq:myIK}{b,d}, that $\myB/\myR^2$ is a key parameter.  We therefore consider separately the cases $\myB\ll \myR^2<1$ and $\myR^2\ll \myB\ll1$. We find the leading-order results
\begin{equation}\label{eq:sol_linbencyl}
\frac{\myd}{\myF} \sim 
\begin{dcases}
\log \frac{e^\gamma}{2\myB^{1/2}}& \text{for $ \myR^2\ll \myB\ll 1$,}\\
\log\frac{1}{\myR}  & \text{for $\myB\ll \myR^2<1$,}
\end{dcases}
\end{equation}
where $\gamma\approx 0.577$ is the Euler--Mascheroni constant \cite[][]{Abramowitz1964}. 

\begin{figure} [ht!]
\centering
\includegraphics[width=13cm]{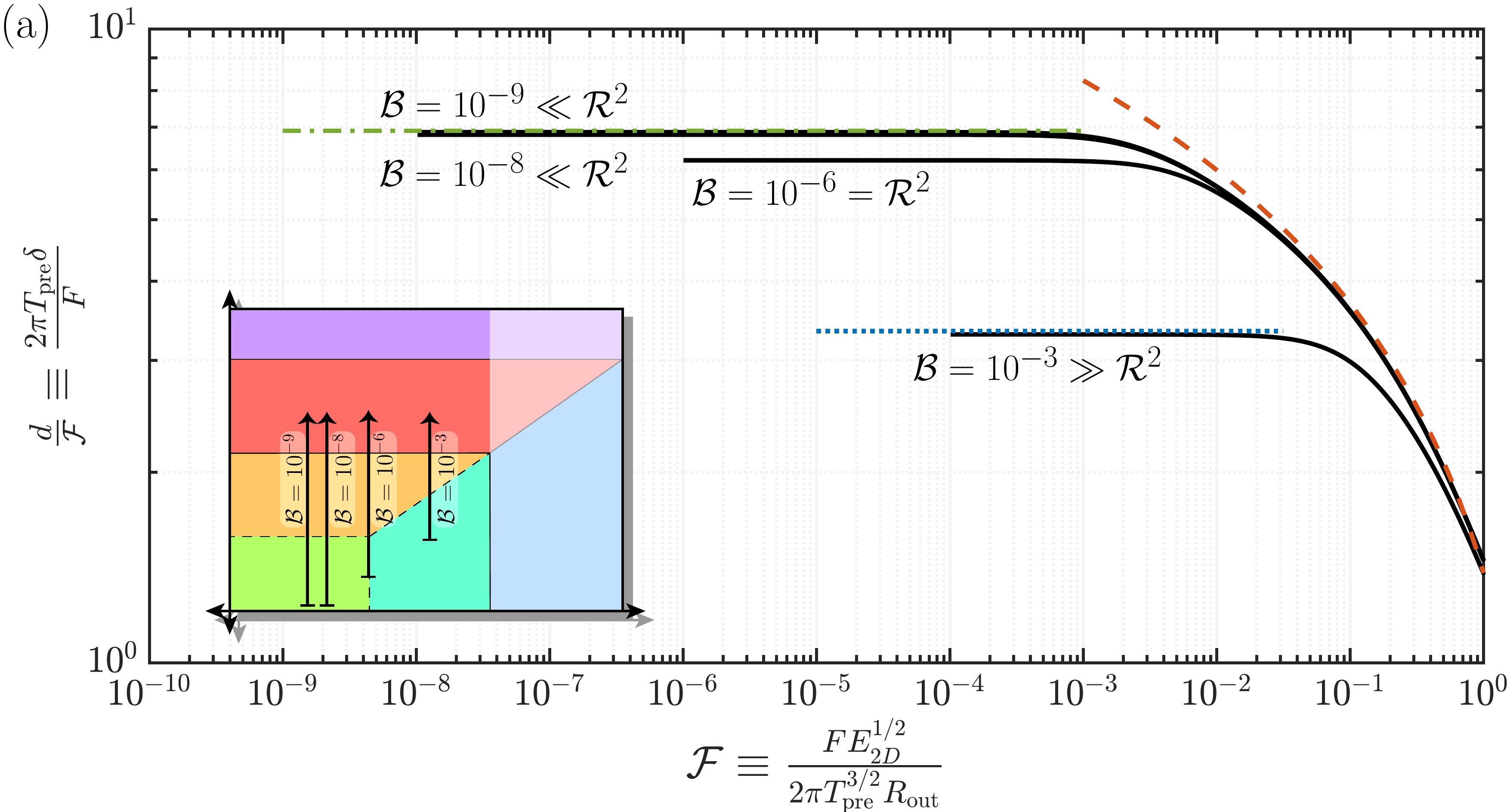}

\vspace*{5pt}

\includegraphics[width=13cm]{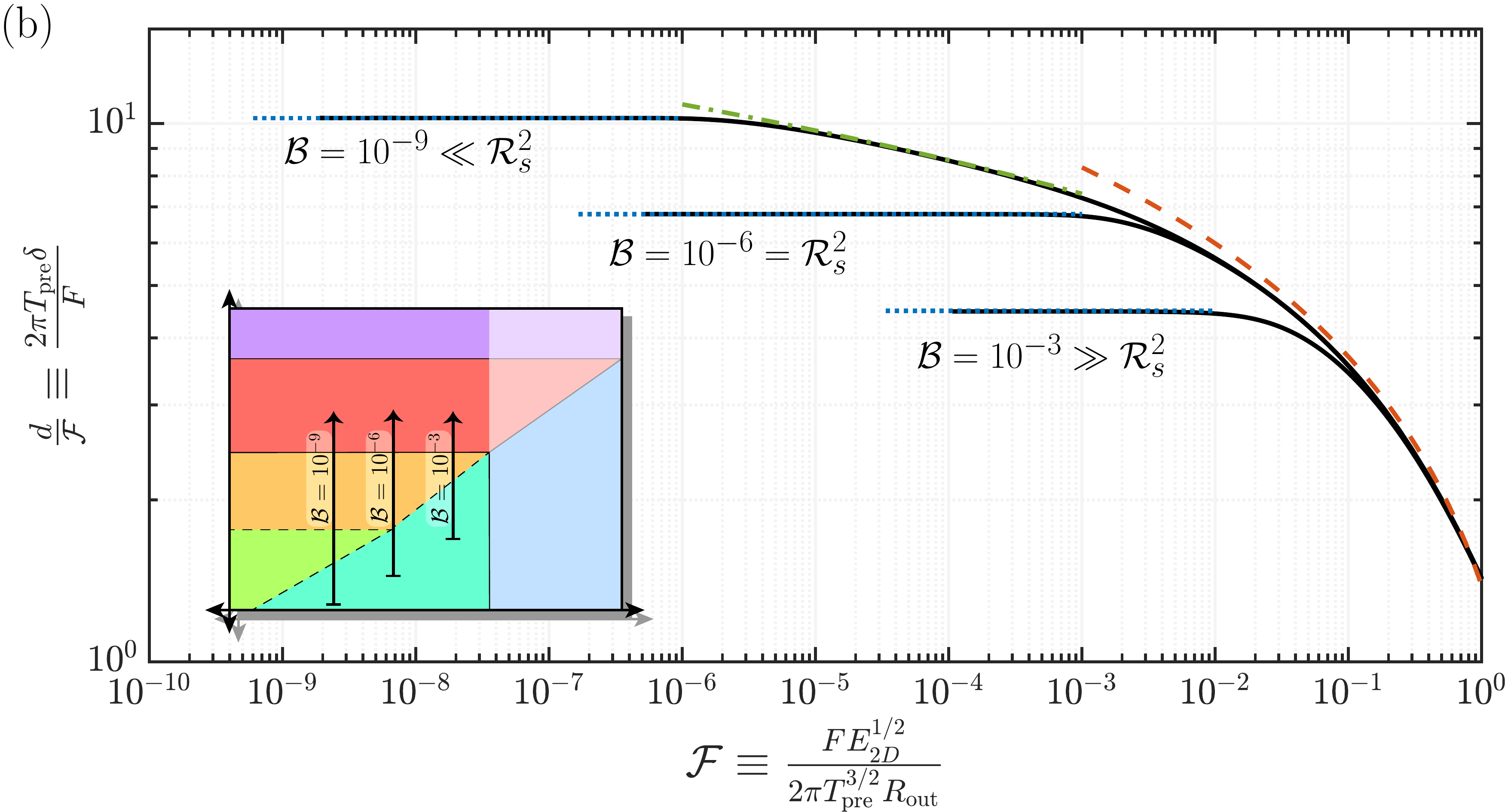}
\caption{The evolution of the linear compliance as a function of imposed force $\myF$ for sheets with different bending stiffnesses ($\myB = 10^{-3}$, $10^{-6}$, $10^{-8}$, $10^{-9}$) and  $\nu=1/3$. Results are shown for (a) a cylindrical indenter ($\myR=10^{-3}$) and (b) a spherical indenter ($\myRs=10^{-3}$). Numerical solutions to the FvK equations are shown as black solid curves while the corresponding asymptotic solutions are shown as broken  curves as follows: eqs.~\eqref{eq:sol_linbencyl}[$\myR^2\ll\myB$]~\&~\eqref{eq:sph_linben}[$\myRs\myF\ll\myB$] in blue (dotted), eqs.~\eqref{eq:sol_linptcyl}[$\myF\ll\myR$]~\&~\eqref{eq:sph_linben}[$\myB\ll\myRs\myF$] in green (dash-dotted), and eqs.~\eqref{eq:sol_linptcyl}[$\myR\ll\myF$]~\&~\eqref{eq:sph_linpt}[$\myRs\ll\myF$] in red (dashed).  The insets show the routes through the relevant $(\myB,\myF)$ parameter space (see the regime diagrams of Fig.~\ref{fig:regime_cylsph}) taken by indentation in each case. \label{fig:fvklinear_ben_sph}}
\end{figure}

The evolution of the compliance with indentation force is shown in Fig.~\ref{fig:fvklinear_ben_sph} for a fixed $\myR$ and four values of $\myB$; we see that, for small $\myF$,  numerical results agree with our asymptotic results. Note, in particular, that the results for $\myB=10^{-8}$ and $\myB=10^{-9}$ are essentially indistinguishable at the scale of the plot, Fig.~\myref{fig:fvklinear_ben_sph}{a}:  in both cases, $\myR^2=10^{-6}\gg\myB$ and so the small-indentation compliance is controlled by the indenter radius, rather than bending stiffness, as predicted by \eqref{eq:sol_linbencyl}. In these cases, the compliance is essentially indistinguishable from that of an ideal membrane with $\myB=0$. 

Overall, for small displacements by a cylindrical indenter ($\myF\ll \max\{\myB^{1/2},\myR\}$), the indentation compliance $\myd/\myF$ is a constant determined by the relative size of the indenter radius $\myR$  and the bending stiffness $\myB^{1/2}$. We note also that the fact that the form of the stiffness in each of the cases in \eqref{eq:sol_linbencyl} is functionally similar --- both are logarithmic --- is not a coincidence: in the bending-dominated case, the sheet is approximately flat over a region of horizontal scale $\myB^{1/2}$, and it is as if the sheet were deformed by a virtual cylindrical indenter of radius $\myR\sim2e^{-\gamma}\myB^{1/2}$. 

While the indenter-dominated solution \eqref{eq:sol_linbencyl}[$\myB\ll\myR^2<1$] has been previously derived \cite[see][for example]{Jennings1995}, we believe that the bending-dominated solution \eqref{eq:sol_linbencyl}[$\myR^2\ll\myB\ll1$] is novel; this describes scenarios when the bending stiffness is small, but cannot be neglected because the indenter size is smaller than the virtual bending-induced  indenter radius, which is therefore the relevant length scale.

\subsection{Small bending stiffnesses $(\myB^{1/2}\ll\max\{\myF,\myR\})$}\label{sec:results_B=0_cyl}

In many experimental setups the dimensionless bending stiffness $\myB$ is small enough that the sheet can be modelled as a thin membrane. In this case, we can simplify the model by taking $\myB \to 0$ in \eqref{eq:fvk_dless1} and dropping the highest-order  boundary  conditions \myeqref{eq:bc_cyl_dless}{b,e} --- ignoring the effects of the edge boundary layers, which occur over a typical length scale $\rho\sim\myB^{1/2}$.

\citet[App.~B]{Vella2017} solved the membrane problem for a point indenter ($\myR=0$); we extend their work by solving the finite cylinder  case (the full derivation can be found in our \ref{app:small_bending}). Ultimately, we obtain the parametric force--displacement relation:
\begin{subequations}\label{eq:B0_full_cyl}
\begin{gather}
\myd = 2\frac{\sinh^{-1}\Phi_1^{1/2}-\sinh^{-1}\Phi_0^{1/2}}{A[\Phi_0,\Phi_1;\myR,\nu]^{1/2}},\qquad \myF = \frac{2}{\myR^{2}}\frac{\Phi_0(1+\Phi_0^{-1})^{-1/2}}{A[\Phi_0,\Phi_1;\myR,\nu]^{3/2}}, \subtag{a,b}\\
A[\Phi_0,\Phi_1;\myR,\nu]\coloneqq \frac{2}{1-\nu}\frac{\Phi_0}{\myR^2}\sqrt{\frac{1+\Phi_1^{-1}}{1+\Phi_0^{-1}}}-\frac{1+\nu}{1-\nu}\Phi_1, \subtag{c}
\end{gather}
which is parametrized by the boundary stresses $\Phi_1\coloneqq A \Psi(1)$ and $\Phi_0\coloneqq A \myR \Psi(\myR)$, these `stresses' must further satisfy the equation
\begin{equation}
\frac{\Phi_1^{3/2}}{\left(1+\Phi_1\right)^{1/2}}- \myR^{-2}\frac{\Phi_0^{3/2}}{(1+\Phi_0)^{1/2}}=\left[\sinh^{-1}\Phi^{1/2} -\left(1+\Phi^{-1}\right)^{-1/2}\right]_{\Phi_0}^{\Phi_1}. \subtag{d}
\end{equation}
\end{subequations}
For a given indentation force $\myF$, equations \myeqref{eq:B0_full_cyl}{b} and \myeqref{eq:B0_full_cyl}{d} form a pair of equations for the two unknowns $\Phi_0$ and $\Phi_1$.  Equation \myeqref{eq:B0_full_cyl}{a} therefore yields an implicit force--displacement relation $\myd(\myF;\myR,\nu)$. In the limit of the inner stress (or equivalently the radius) being taken to zero,  $\Phi_0=\Oh(\myR^{4/3})\to 0$, this system yields the point-indenter result of \citet[eq.~(12)--(14)]{Vella2017}.

To make further analytical progress with a finite indenter radius, we  concentrate on the two asymptotic limits $\myF\to 0$ (\S\ref{sec:moderate_forces}) and $\myF\to\infty$ (\S\ref{sec:large_forces}). This allows us to determine explicit forms that are formally valid only in these limits, but that might be expected to apply more broadly. 

\subsubsection{Moderate indentation forces $(\myF\ll 1)$}\label{sec:moderate_forces}

The small indentation limit $\myF\to0$ is directly equivalent to taking $A[\Phi_0,\Phi_1]\to\infty$; consequently we find that $A\sim\Phi_1=4/\myF^{2} + \Oh(1)$ as $\myF\to0$. Inserting these into \eqref{eq:B0_full_cyl} gives the implicit force--displacement relation:
\begin{subequations}\label{eq:stiff_Fsmall}
\begin{equation}
\frac{\myd}{\myF}=\log\frac{4}{\myF}-\sinh^{-1} \Phi_0^{1/2} + \Oh(\myF^{2}),
\end{equation}
where $\Phi_0$ is given by 
\begin{equation}
\myR^{-2}\frac{\Phi_0^{3/2}}{(1+\Phi_0)^{1/2}}= \frac{4}{\myF^2}+\Oh(1).
\end{equation}
\end{subequations}
The parameter $\Phi_0$ can be eliminated from \eqref{eq:stiff_Fsmall} to give an explicit equation for $\myd$ in terms of $\myF$; however, the result is complicated and does not readily reveal the limiting behaviours. Instead we consider the sub-cases $\myF\ll \myR$ and $\myF\gg \myR$, which correspond to taking $\Phi_0\to \infty$ and $\Phi_0 \to 0$, respectively. 

At leading-order we find
\begin{equation}\label{eq:sol_linptcyl}
\dfrac{\myd}{\myF}\sim \begin{dcases}\log\frac{1}{\myR} &\text{for $\myF\ll \myR < 1$}, \\
\log\frac{4}{\myF} &\text{for $\myR\ll \myF \ll 1$}.
\end{dcases}
\end{equation}
The very small displacement solution in \eqref{eq:sol_linptcyl} (found when $\myF\ll\myR$) matches precisely with our earlier solution from \eqref{eq:sol_linbencyl} with $\myB\ll\myR^2$. Moreover, the moderate displacement solution in \eqref{eq:sol_linptcyl} (found when $\myR\ll\myF\ll 1$) is that presented by \cite{Vella2017} for a point indenter, and is shown as the red-dashed curve in Fig.~\ref{fig:fvklinear_ben_sph}. In summary, for a cylinder, as the indentation force increases, the indentation compliance $\myd/\myF$ evolves from being a constant (controlled by the indenter radius) to a logarithmic behaviour (controlled by an $\myF$-dependent stretching-induced radius).

\subsubsection{Large indentation forces $(\myF\gg1)$}\label{sec:large_forces}

The large indentation limit $\myF\to\infty$ is  equivalent to taking $A[\Phi_0,\Phi_1]/\Phi_1\to 0$; consequently we find that $A= \Oh(\myF^{-2/3})$ and $\Phi_1=\Oh(1)$ as $\myF\to \infty$. Inserting these into \eqref{eq:B0_full_cyl} gives the force--displacement relation:
\begin{subequations}\label{eq:cubic_cyl_full}
\begin{equation}
\frac{\myd}{\myF^{1/3}}  = \frac{2}{(1+\nu)^{1/3}}\frac{\sinh^{-1}\Phi_1^{1/2}-\sinh^{-1}\Phi_0^{1/2}}{\Phi_1^{1/2}\left(1+\Phi_1\right)^{-1/6}}+\Oh(\myF^{-2/3}),
\end{equation}
where $\Phi_1$ and $\Phi_0$ are the solutions to
\begin{gather}
 \myR^{-2}\frac{\Phi_0^{3/2}}{\left(1+\Phi_0\right)^{1/2}}=\frac{(1+\nu)}{2}\frac{\Phi_1^{3/2}}{\left(1+\Phi_1\right)^{1/2}}+\Oh(\myF^{-2/3}),\\
\frac{1-\nu}{2}\Phi_1=\frac{\left(1+\Phi_1\right)^{1/2}}{\Phi_1^{1/2}} \left[\sinh^{-1}\Phi^{1/2} -\left(1+\Phi^{-1}\right)^{-1/2}\right]_{\Phi_0}^{\Phi_1}+\Oh(\myF^{-2/3}).
\end{gather}
\end{subequations}
The expression in (\ref{eq:cubic_cyl_full}a) corresponds to a constant cubic compliance: $d/\myF^{1/3}$ is a function of the indenter size $\myR$ and Poisson's ratio $\nu$ alone. 

\begin{figure}[ht!]
\centering
\includegraphics[width=13cm]{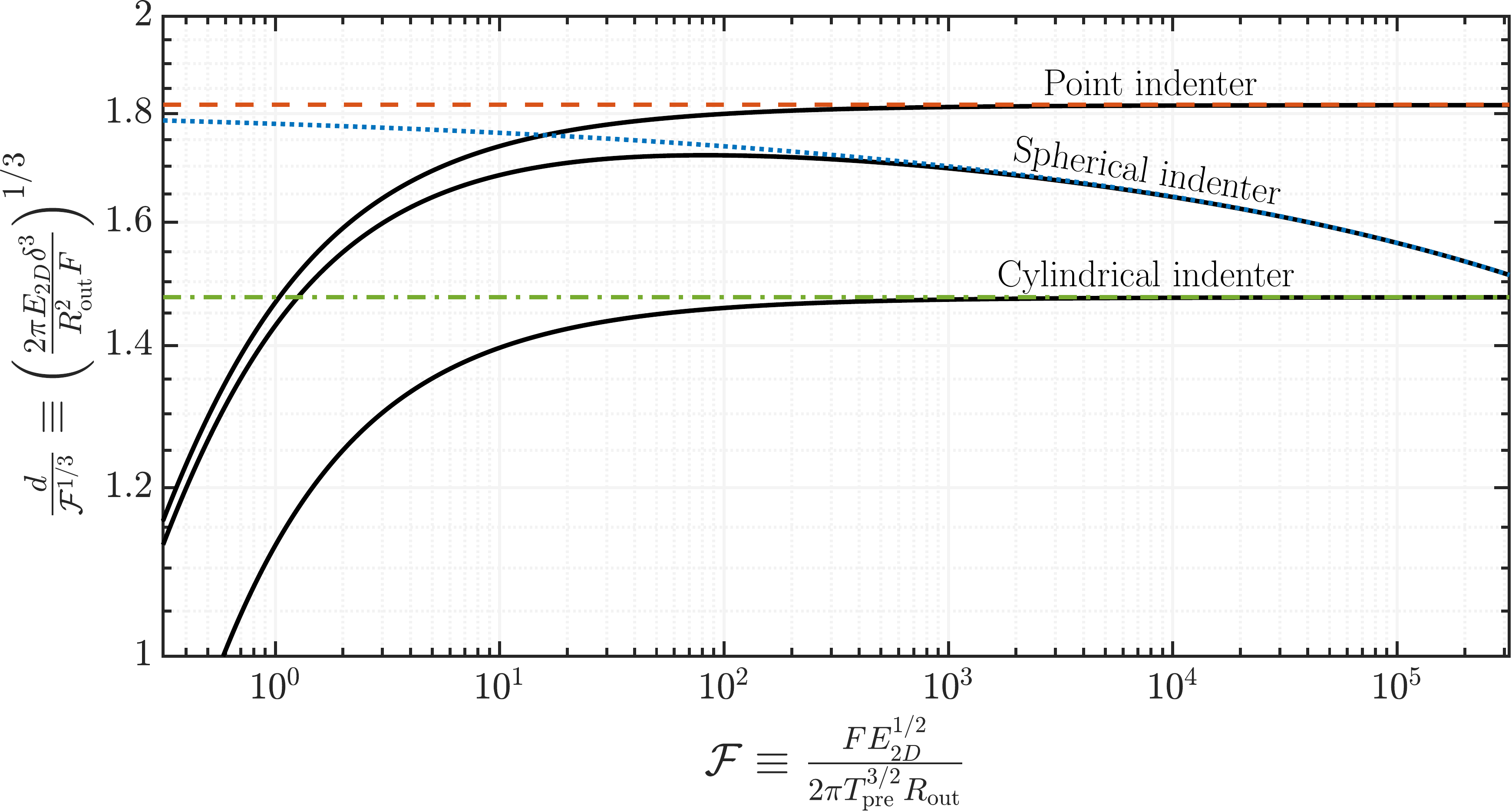}
\caption{The cubic compliance associated with large indentations of a linear-elastic membrane ($\nu=1/3$, $\myB=0$) for indenters of different type: results are shown for point ($\myR\to 0$), cylindrical ($\myR=10^{-1}$), and spherical-capped ($\myRs=10^{-3}$) indenters. Numerical solutions of the FvK equations are shown as solid black curves while the corresponding asymptotic solutions are shown as  broken curves as follows: eq.~\eqref{eq:cubic_cyl}[$\myR=0$] in red (dashed), eq.~\eqref{eq:cubic_sph} in blue (dotted), and eq.~\eqref{eq:cubic_cyl} in green (dash-dotted). Observe the difference made by geometry: results for a spherical indenter never reach the constant cubic compliance regime $\myd/\myF^{1/3}=\mathrm{cst}$, while even for cylindrical indenters the finite indenter size may play a significant role.}
\label{fig:fvkcubic_cyl_sph}
\end{figure}

Experimentally, the indenter is often orders of magnitude smaller than the sheet clamping radius, so that $\myR\ll1$ \citep{Lee2008}; this limit corresponds to $\Phi_0=\Oh(\myR^{4/3})\to 0$, and so the cylindrical result \eqref{eq:cubic_cyl_full} simplifies to
\begin{subequations}\label{eq:cubic_pt_full}
\begin{equation}
\frac{\myd}{\myF^{1/3}}  = \frac{2}{(1+\nu)^{1/3}}\frac{\sinh^{-1}\Phi_1^{1/2}}{\Phi_1^{1/2}\left(1+\Phi_1\right)^{-1/6}}-(2\myR)^{2/3}+\Oh(\myF^{-2/3},\myR^2),
\end{equation}
where $\Phi_1$ is the solution to
\begin{equation}
1+\frac{1-\nu}{2}\Phi_1=\frac{\left(1+\Phi_1\right)^{1/2}}{\Phi_1^{1/2}} \sinh^{-1}\Phi_1^{1/2}+\Oh(\myF^{-2/3},\myR^2).
\end{equation}
\end{subequations}
This is the point-indenter result from \cite{Vella2017} with an additional $(2\myR)^{2/3}$ term to account for the small (but finite) indenter size. We find that the cubic compliance
\begin{equation}
\label{eq:cubic_cyl}
\frac{d}{\myF^{1/3}} \sim q_c[\nu,\myR]= q_p[\nu]-(2\myR)^{2/3} + \mathcal{O}(\myR^2),
\end{equation}
where $q_c[\nu,\myR]$ and $q_p[\nu]$ are given by the leading-order equations in \eqref{eq:cubic_cyl_full} and \eqref{eq:cubic_pt_full} respectively. Thus, for large indentations by a cylindrical indenter ($\myB^{1/2}\ll1\ll\myF$), the cubic compliance $\myd/\myF^{1/3}$ is a constant controlled by the sheet's Poisson's ratio $\nu$ and indenter radius $\myR$. An analogous result was found by \citet[eq.~(48)]{Vella2017}, for the case of a no-slip indenter: $d/\myF^{1/3} \sim \alpha_0(\nu)^{-1/3}-[8\myR^2/(1+\nu)]^{1/3}$ where $\alpha_0(\nu)^{-1/3}\equiv q_p[\nu]$.  Figure~\ref{fig:fvkcubic_cyl_sph} shows the comparison between numerical simulations, the prediction for a point indenter $\myR=0$, and the expression \eqref{eq:cubic_cyl} for $\myR=0.1$. We see that the effect of finite indenter size is non-negligible in the large force limit.

The asymptotic predictions  \eqref{eq:sol_linbencyl}, \eqref{eq:sol_linptcyl}, and \eqref{eq:cubic_cyl} are presented in the regime diagram Fig.~\myref{fig:regime_cylsph}{a}. We now move on from the case of a cylindrical indenter to one with a  spherical tip.

\section{Spherical indentation of a linearly elastic sheet}\label{sec:sphere}

Having considered in some detail the simplest case of a finite cylindrical indenter, we now move on to a case of more experimental relevance: an indenter with a spherical tip  (\eg~\citealt{Bhatia1968, Jennings1995, Begley2004, Lee2008, LopezPolin2017}). The key difference between this case and  the cylindrical indenter already considered is that the radial position of the edge of contact, $\Rin$, is initially unknown and evolves with the indentation depth, $\delta$, as the sheet wraps more of the indenter. 

\subsection{F\"oppl--von K\'arm\'an formulation}

The F\"oppl--von K\'arm\'an formulation for a spherical-capped indenter is the same as the cylindrical indenter formulation in \S\ref{sec:cylinder_fvk} but with inner boundary conditions modified to account for the new indenter geometry. For consistency with the assumptions inherent in the FvK equations, we approximate the tip as a parabola. The analogue of the boundary conditions \myeqref{eq:bc_cyl_dless}{d--f} therefore take the dimensionless form
\begin{subequations}\label{eq:bc_sph_dless}\begin{align}
 Z(\myR) = -\myd +\frac{\myR^2}{2\myRs}, \qquad Z'(\myR) &=\frac{\myR}{\myRs},\qquad Z''(\myR) =\frac{1}{\myRs},\subtag{a--c} \\
 \Psi'(\myR)- \frac{\Psi(\myR)}{\myR} &= -\frac{\myR^2}{8\myRs^2}.\subtag{d}
\end{align}\end{subequations}
Here \myeqref{eq:bc_sph_dless}{a--c} express the geometric conditions of continuous vertical displacement, slope, and curvature, while \myeqref{eq:bc_sph_dless}{d} comes from an in-plane force balance, since within the contact region it may easily be shown that  $\Psi(r)+\rho^3/16\myRs^2\propto \rho$. We again emphasize that the point of contact $\myR$ is unknown here and must be determined as part of the solution, explaining why we require an extra boundary condition \myeqref{eq:bc_sph_dless}{c} compared to the cylindrical case. 

The system of equations~\eqref{eq:fvk_dless} subject to \myeqref{eq:bc_cyl_dless}{a--c} and \eqref{eq:bc_sph_dless} can be solved for given parameters $\nu$, $\myB$, $\myF$, and $\myRs$ by using a standard numerical integrator with unknown $\myR$. To facilitate this computation,  it is  more convenient to fix the value of $\myR$ and solve for an unknown $\myF$ instead. By doing so, one avoids the problems associated with an unknown domain size. 

To make analytical progress we  apply the same asymptotic simplifications as in the cylindrical case: (i) linearize around the pre-stretched base state [\ie~take $\Psi(\rho)\sim \rho +\tilde{\Psi}(\rho)$ and $Z(\rho)\sim 0+\tilde{Z}$ for $\tilde{\Psi}, \tilde{Z}\ll 1$]; or (ii) consider the membrane theory limit [\ie~take $\myB\to 0$ and drop the boundary conditions \myeqref{eq:bc_cyl_dless}{b} and \myeqref{eq:bc_sph_dless}{c}]. The analysis of these asymptotic limits are  analogous to the cylindrical indenter case presented in \S\ref{sec:results_B>0_cyl}~\&~\S\ref{sec:results_B=0_cyl}, but with the added detail that $\myR$ is unknown; we shall only present the final results below (the full analysis can be found in \ref{app:small_forces}, for small indentation forces, and \ref{app:small_bending}, for vanishing bending stiffnesses).

\subsection{Small indentation forces $(\myF\ll\max\{\myB^{1/2},\myRs\})$}\label{sec:results_B>0_sph}

Linearizing around the pre-tensed state, we find that
\begin{subequations}\label{eq:sph_linben_full}
\begin{equation}\label{eq:sph_linben}
\frac{\myd}{\myF} \sim 
\begin{dcases}
\log \frac{e^\gamma}{2\myB^{1/2}}& \text{for $ \myRs\myF \ll \myB\ll 1$,}\\
\log\sqrt{\frac{e}{\myRs\myF}} & \text{for $\myB\ll \myRs\myF<1$,}
\end{dcases} 
\end{equation}
with
\begin{equation}\label{eq:sph_linben_R}
\myR^{2} \sim 
\begin{dcases}
4\myB e^{-2\gamma}\exp\left[-\frac{4\myB}{\myRs\myF}\right]& \text{for $ \myRs\myF\ll \myB\ll 1$,}\\
\myRs\myF& \text{for $\myB\ll \myRs\myF<1$,}
\end{dcases} 
\end{equation}\end{subequations}
where $\gamma\approx 0.577$ is again the Euler--Mascheroni constant. Thus, for small displacements by a spherical-capped indenter ($\myF\ll \max\{\myB^{1/2}, \myRs\}$), the indentation compliance $\myd/\myF$ evolves from being a constant (controlled by the bending-induced radius that was discussed in the cylindrical indenter problem) to a logarithmic behaviour (controlled by the radius of contact, which in turn depends on the force). This evolution is shown in Fig.~\ref{fig:fvklinear_ben_sph} and confirms that the numerical results reproduce the expected asymptotic results in the relevant limits.

The asymptotic compliance \myeqref{eq:sph_linben_full}{a} is directly equivalent to that for a  cylindrical indenter, \ie~\eqref{eq:sol_linbencyl}, and can be recovered by accounting for the extra indentation depth due to the sphere geometry [$\myd\mapsto \myd+\myR^2/2\myRs$, \ie~compare \myeqref{eq:bc_cyl_dless}{d} and \myeqref{eq:bc_sph_dless}{a}] and inserting the contact radius expression \eqref{eq:sph_linben_R}. It should be noted that the indenter-dominated solution \eqref{eq:sph_linben_full}[$\myB\ll\myRs\myF$] has also been derived by  \citet[eq~(43b)]{Bhatia1968}, from analytical solutions to small-rotation Reissner theory (analogous to FvK), and \citet[eq.~(14)]{Norouzi2006}, by  minimizing the energy of the constant-tension problem. 

\subsection{Moderate indentation forces $(\min\{\myB^{1/2},\myB/\myRs\}\ll\myF\ll 1)$}\label{sec:results_B=0_sph1}
Taking the membrane theory limit and assuming small indentation forces ($\myF\ll 1$) we find that
\begin{subequations}\label{eq:sph_linpt_full}
\begin{equation}\label{eq:sph_linpt}
\dfrac{\myd}{\myF}\sim \begin{dcases}\log\sqrt{\frac{e}{\myRs\myF}}  &\text{for $\myF\ll \myRs \ll 1$}, \\
\log\frac{4}{\myF} &\text{for $\myRs\ll \myF \ll 1$},
\end{dcases}
\end{equation}
with
\begin{equation}\label{eq:sph_linpt_R}
\myR^{2} \sim 
\begin{dcases}
\myRs\myF& \text{for $ \myF\ll \myRs\ll 1$,}\\
4(\sqrt{2}-1)\sqrt{\myRs^{3}\myF}& \text{for $\myRs\ll \myF\ll 1$.}
\end{dcases}
\end{equation}\end{subequations}
Thus, for moderate indentations by a spherical-capped indenter ($\min\{\myB^{1/2},\myB/\myRs\}\ll\myF\ll 1$), the indentation stiffness $\myF/\myd$ evolves from one logarithmic behaviour (controlled by the radius of contact) to another (controlled by a stretching-induced radius); these asymptotic results are also shown in Fig.~\ref{fig:fvklinear_ben_sph}.

Solution~\eqref{eq:sph_linpt_full} is directly equivalent to the cylindrical result \eqref{eq:sol_linptcyl}, and can again be obtained by accounting for the extra indentation depth due to the sphere geometry [$\myd\mapsto \myd+\myR^2/2\myRs$, \ie~compare \myeqref{eq:bc_cyl_dless}{d} and \myeqref{eq:bc_sph_dless}{a}] and inserting the contact radius expression \eqref{eq:sph_linpt_R}.

\subsection{Large indentation forces $(\myF\gg1)$}
\label{sec:results_B=0_sph2}
Taking the membrane theory limit and assuming large indentation forces ($\myF\gg 1$), we find that
\begin{subequations} \label{eq:cubic_sph_full}
\begin{equation}\label{eq:cubic_sph}
\frac{d}{\myF^{1/3}} \sim q_s[\nu,\myRs^3\myF]= q_p[\nu]-2(2-\sqrt{2})(\myRs^3\myF)^{1/6} + \mathcal{O}(\sqrt{\myRs^{3}\myF}),
\end{equation}
with
\begin{equation}\label{eq:cubic_sph_R}
\myR^2 \sim 4(\sqrt{2}-1)\sqrt{\myRs^{3}\myF}.
\end{equation}\end{subequations} 
Note that this result suggests a different effect of the geometry of the indenter, and the associated change in contact radius, than that proposed by \cite{Jia2020} based on a linear fitting procedure. We also emphasize that this result is different from the results presented by \cite{Begley2004} who neglect azimuthal strain and axial force balance to facilitate an approximate solution of the governing equations; in our notation this leads them to identify a constant cubic compliance, dependent on $\Rs/\Rout$. Instead, our asymptotic analysis shows that the cubic compliance is \emph{not} constant but has a correction at  $\Oh( \myRs^{1/2}\myF^{1/6})$ that makes it weakly dependent on the force and indenter radius. 

 Above, we have implicitly assumed that $\myRs^{3}\myF\ll1$, which must be true for the assumption of small slopes to hold ($\myR\ll\myRs\ll1$). Thus, for large indentations by a spherical indenter ($\myF\gg 1$), the cubic compliance $\myd/\myF^{1/3}$ is controlled by the sheet's Poisson's ratio $\nu$ and contact radius $\myR\sim(\myRs^3\myF)^{1/4}$. This is shown in Fig.~\ref{fig:fvkcubic_cyl_sph}, and demonstrates that in the spherical case the cubic compliance never saturates at a constant value (while it does for a cylindrical indenter). Nevertheless, the numerically-observed behaviour of the cubic compliance is in good agreement with the asymptotic prediction \eqref{eq:cubic_sph_full}.

To our knowledge, the solution \eqref{eq:cubic_sph_full}  is new. Moreover, while it is functionally similar to the corresponding cylindrical result \eqref{eq:cubic_cyl}, it cannot be obtained by accounting for the extra indentation depth [$\myd\mapsto \myd+\myR^2/2\myRs$] and  inserting the contact radius expression \eqref{eq:cubic_sph_R}, as was possible for small indentation depths. This difference is because the stress in the membrane that is in contact with the indenter is different between the two models --- compare \myeqref{eq:bc_cyl_dless}{f} and \myeqref{eq:bc_sph_dless}{d}. The asymptotic results \eqref{eq:sph_linben}, \eqref{eq:sph_linpt}, and \eqref{eq:cubic_sph} are presented in the regime diagram Fig.~\myref{fig:regime_cylsph}{b}. 

\subsection{Numerical results for a spherical indenter}

\begin{table}[ht!]
\centering
\resizebox{\textwidth}{!}{%
\begin{tabular}{l | c | c | c | c | c }
Reference & Material & $\dfrac{\Rs}{\Rout}$  &  $F_{\max}$ [$\mathrm{n N}$]& $  \dfrac{\Rout\Tpre^{3/2}}{\Y^{1/2}}$ [$\mathrm{nN}$] &  $\Y\Rs$ [$\mathrm{n N}$]  \\[1.2em]
\hline\hline & & & & & \\[-1em]
\cite{Lee2008} &Graphene & $0.02$ -- $0.06$ & $ 1200$ -- $2900$&  $  0.5$ -- $25.5$ & $5600$ -- $9400$   \\
\cite{Song2010}  &h-BN& $0.05$ & $ 200$& --- & $1100$ -- $25500$  \\
\cite{Bertolazzi2011}&$\mathrm{MoS_2}$ & $0.05$ & $200$& $ 0.05$ -- $0.5$ &$2200$   \\
\cite{Lee2013}  &Graphene & $0.03$ -- $0.08$ & $ 2000$& $ 0.9$ -- $1.3$  &$8600$ -- $12500$  \\
\cite{LopezPolin2017} &Graphene & $0.02$ -- $0.12$ & $ 1200$ -- $2100$ &   $  0.2$ -- $80$ & $ 1500$ -- $4000$   \\
\cite{Harats2020} &$\mathrm{WS_2}$& $0.02$ -- $0.10$ & $  300$ --  $900$& $  0.7$ -- $4.3$ & $ 8500$  
\end{tabular}
}
\caption{Typical experimental values of the  parameters relevant to our model as determined in previous indentation experiments on various two-dimensional materials. (h-BN is hexagonal boron nitride, $\mathrm{MoS_2}$ is molybdenum disulphide and $\mathrm{WS_2}$ is tungsten disulphide.) Values of $\Rs/\Rout$ are used to inform the parameters used in the numerical results shown in Fig.~\ref{fig:FvKSpherical}, while typical values of the maximum indentation force (denoted $F_{\max}$) reached experimentally for graphene are indicated by the shaded region in Fig.~\ref{fig:FvKSpherical}. The values in the last two columns are pertinent to the discussion in \S\ref{sec:MeasureNonHookean}, and are calculated using the reported values of $\Y$ and $\Tpre$ in the corresponding reference. \label{tab:experiments}}
\end{table}

\begin{figure}[ht!]
\centering
\includegraphics[width=12cm]{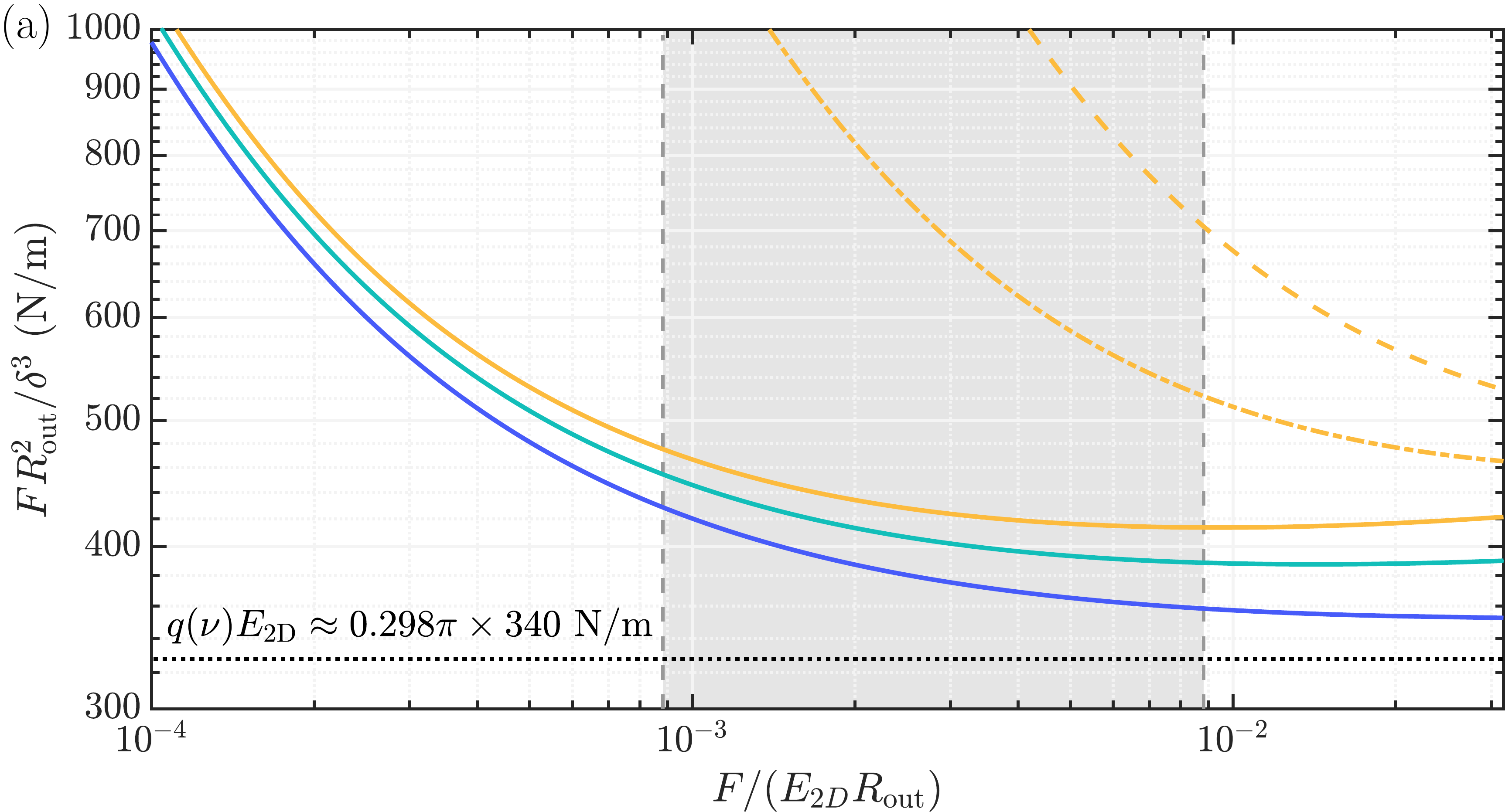}

\vspace*{5pt}

\includegraphics[width=12cm]{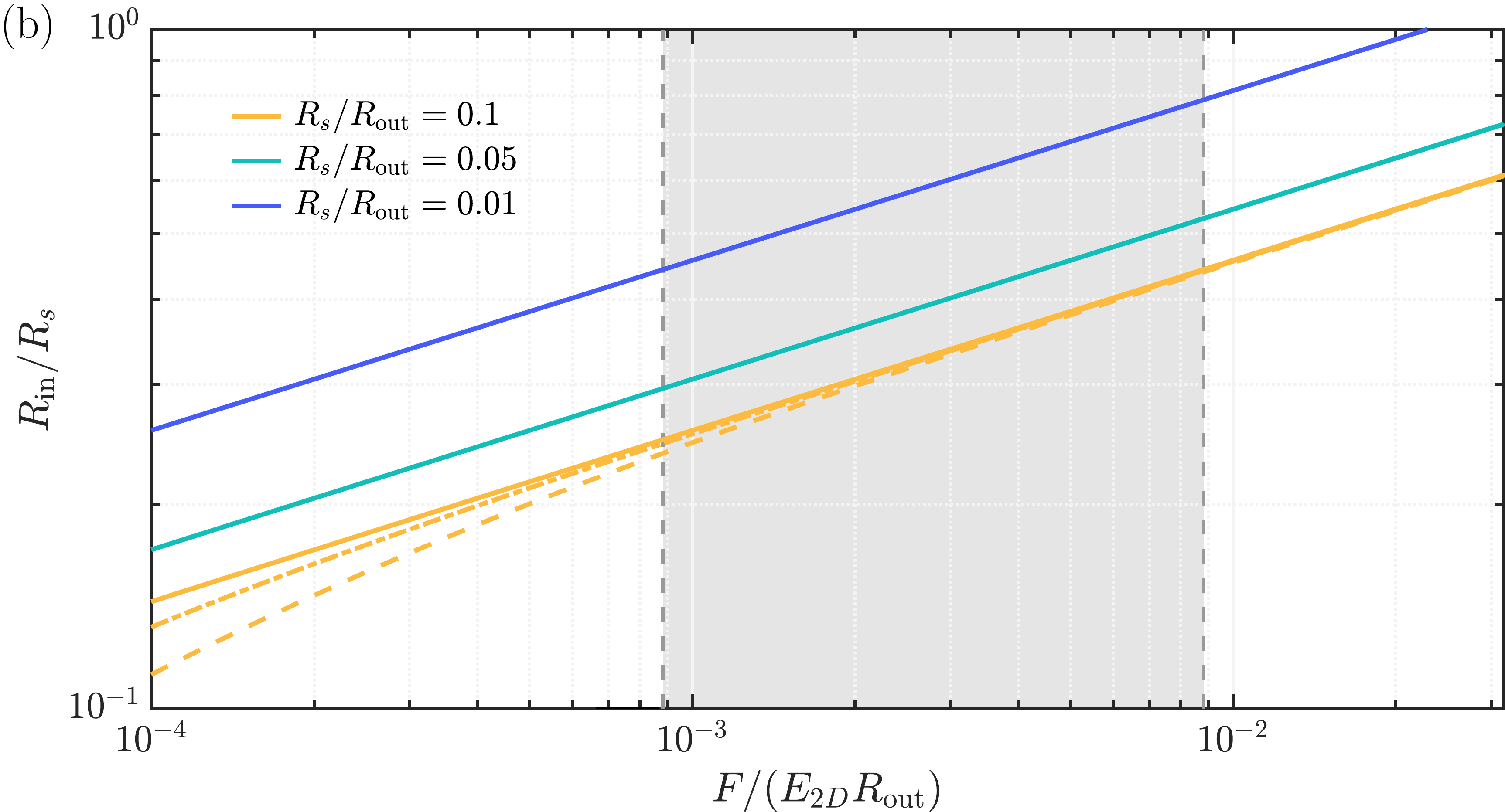}
\caption{Numerical results obtained from the F\"{o}ppl--von K\'{a}rm\'{a}n model with a spherical-capped indenter show the nonlinear effects associated with indenter geometry. (a) The instantaneous estimate of $q(\nu)\Y(=F\Rout^2/\delta^3)$ is plotted as a function of $F/(\Y\Rout)$ for different values of $\Rs/\Rout$ [indicated by colour: $\Rs/\Rout=0.1$ (yellow), $\Rs/\Rout=0.05$ (green) and $\Rs/\Rout=0.01$ (blue)] and pre-tension $\Tpre/\Y$ [indicated by line style: $\Tpre/\Y=10^{-3}$ (solid curves), $\Tpre/\Y=5\times10^{-3}$ (dash-dotted curves) and $\Tpre/\Y=10^{-2}$ (dashed curves)]. The prediction of \cite{Schwerin1929} for a point indenter with zero pre-tension is shown by the horizontal black dotted line. The typical range of experimental indentation forces used when fitting for the stretching modulus of graphene ($ 300~\mathrm{nN}\lesssim F\lesssim 3000~\mathrm{nN}$, as described in Table~\ref{tab:experiments}) is indicated by the shaded region under the assumption that $\Y\Rout=3.4\times10^{-4}\mathrm{~N}$. (b) The evolution of the radius at the edge of contact, $\Rin/\Rs$, is plotted as a function of force, with the same key as in (a)  illustrating the varying pre-tension and sphere size. In both plots, true values  $\nu=0.165$ and $\Y=340\mathrm{~N/m}$   are assumed to present the numerical results in dimensional form. 
\label{fig:FvKSpherical}}
\end{figure}

Figure \ref{fig:FvKSpherical} shows numerical results for the force--indentation relationship and the contact radius obtained from our FvK model with a spherical indenter. A common means of plotting experimental data is to plot the instantaneous estimate of $q(\nu)\Y$ from the point indenter model,  defined as $q(\nu)\Y \coloneqq  F\Rout^2/\delta^3$ with $q(\nu)\coloneqq 2\pi q_p[\nu]^{-3}$ in our notation, as a function of applied force. Figure~\myref{fig:FvKSpherical}{a} mimics this by plotting $q(\nu)\Y$ as a function of $F/(\Y\Rout)$. Results are shown for a range of assumed pre-tensions and sphere radii, with the typical maximal applied loads  used in previous AFM indentation experiments indicated (see Table~\ref{tab:experiments} for details).

These numerical results show three  particular features. Firstly, increasing the pre-tension may have the (undesired) effect of preventing measurements from approaching the horizontal asymptote before the maximum indentation force that can be applied is reached. Secondly, results with  larger indenter radii of curvature, $\Rs/\Rout$, enhance the effect of the geometric nonlinearity, again pushing results further from the ideal point indenter solution of \cite{Schwerin1929} and making it more difficult to infer the true value of $\Y$. Thirdly, it seems that the fraction of the indenter that is wrapped by the membrane is relatively large, with $\Rin/\Rs$ lying in the interval $0.3\lesssim\Rin/\Rs\lesssim0.7$ in the regime of experimental interest.

More importantly, however, the results of Fig.~\ref{fig:FvKSpherical} show that the cubic  compliance does not asymptote to a constant for spherical indenters with realistic material parameters, as assumed previously \cite[see Figure 6 of][for example]{Begley2004}. Instead, the cubic compliance asymptotically decreases with increasing $\myF$, as described by  \eqref{eq:cubic_sph} at large indentation forces. A decrease in cubic compliance corresponds to an increase in the instantaneous estimate of $\Y\propto F/\delta^3$, as seen in Fig.~\myref{fig:FvKSpherical}{a}; it therefore seems plausible that such non-constant behaviour might be interpreted experimentally as a nonlinear material effect, either softening or stiffening. Before discussing this possibility further, we turn now to consider such material nonlinearities.

\section{Nonlinear elastic materials}\label{sec:nonlinear_model}

Thus far, through the application of the F\"oppl--von K\'arm\'an equations, we have retained the leading-order geometrical nonlinearities associated with deformation, but have assumed that the material response remains Hookean: we have neglected any effect of nonlinear constitutive response. A simple energetic scaling  (\S\ref{sec:moderate_strains}) shows that this assumption is valid provided that both $\myT\ll1$ and $\myF\ll \myR\myT^{-3/2}$, where $\myT$ is the dimensionless pre-tension defined in \eqref{eq:myT}. Many recent experiments have shown behaviour different to that expected on the basis of the FvK model of point indentation, and concluded that they are probing the nonlinear mechanical response of graphene (\eg~\citealt{Lee2008,LopezPolin2017}). However, we have also seen that the geometry of a spherical indenter can give behaviour that differs from the usual cubic response (or constant cubic compliance) expected from the FvK equations. The question, therefore, is what happens when the indentation advances beyond the small-strain limit and how should one distinguish this regime from the geometrically nonlinear effects associated with  indenter shape?  

We introduce a model that allows for the possibility of large slopes and material strains.   Specifically, we use the work of \cite{Green1960}, who derived a generalized model for the large deformations of an elastic membrane.  This formulation allows the membrane to have a stress-strain relationship that is nonlinear (and hence the solid is non-Hookean); we constrain the elastic constants introduced to recover the stretching modulus, $\Y$,  at small strains, and do not refer to an instantaneous effective stretching modulus at finite strains. The formulation presented below follows similar work by \citet{Yang1970,Long2010, Pearce2011,Laprade2013} all of whom built upon \citeauthor*{Green1960}'s formulation.

We  concentrate on the specific case of the indentation by  a perfect-slip, spherical-capped wedge (with radius of curvature $\Rs$ and wedge angle $2\epsilon\ll1$, as shown in Fig.~\ref{fig:nl_sketch}). This type of indenter is commonly used in experiments (\eg~\citealt{Lee2008,LopezPolin2017}) and reduces to the spherical-capped indenter used in  \S\ref{sec:sphere} provided the indentation depth is sufficiently small. (Hence the results of this section should deviate from those of \S\ref{sec:sphere} only at large indentation depths.) Here, we also ignore the effects of the bending stiffness ($\myB=0$), since we are ultimately interested in the stretching dominant limit.

\subsection{Governing ODEs}

\begin{figure}[ht!]
\centering
\includegraphics[width=13cm]{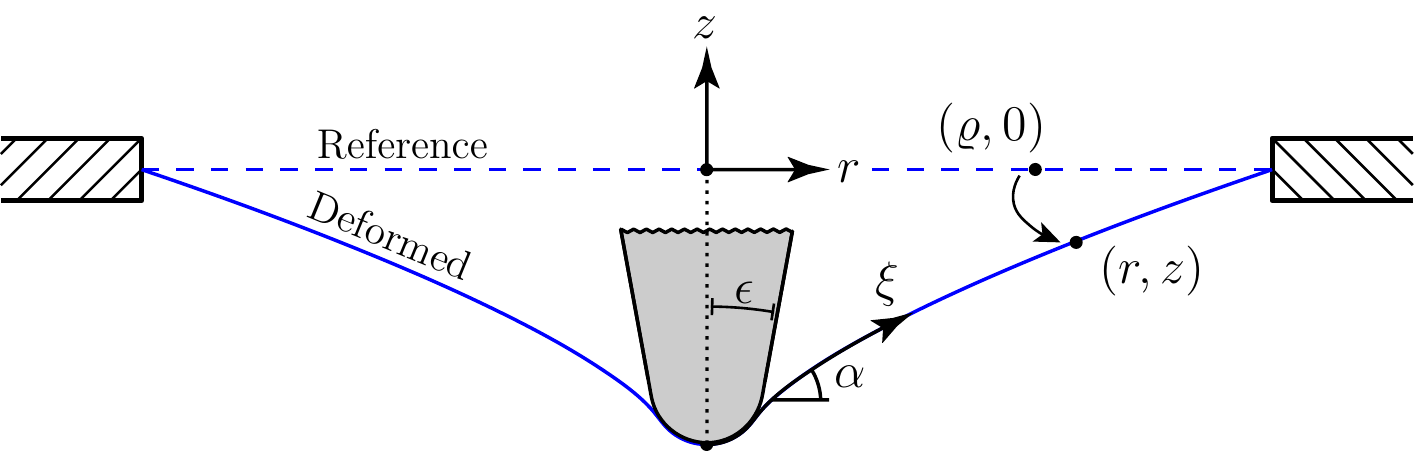}
\caption{Cross-sectional sketch of the indentation of a clamped membrane showing the intrinsic and cylindrical coordinates $(\xi,\alpha)$ and $(r,z)$, respectively. The reference configuration of the membrane is illustrated by the dotted line and the deformed configuration by the solid curve. Note the different variables used to describe the reference and deformed configurations. }
\label{fig:nl_sketch}
\end{figure}

To allow for large  rotations of the sheet, it is useful to introduce intrinsic coordinates (the radial arc-length $\xi$ and angle of rotation $\alpha$, which is measured with respect to the radial-$r$-axis). We then have the geometrical conditions
\begin{subequations}\label{eq:nl_reqs}
\begin{equation}
\frac{\de z}{\de\xi} = \sin\alpha \quad \text{and} \quad \frac{\de r}{\de\xi} = \cos\alpha, \subtag{a,b}
\end{equation}
where we recall that $0\leq r\leq \Rout$  and $z$ are the radial and vertical coordinates of the sheet, respectively.  These variables are sketched in Fig.~\ref{fig:nl_sketch}. 

The membrane is then split into two regions: the region in which the membrane contacts the tip and a non-contacting region. In  the contacting region (which occupies $0\leq r\leq \Rin$ with $\Rin$ as yet unknown) we require 
\begin{equation}\label{eq:sphere_wedge}
\sin\alpha = \begin{dcases}
\frac{r}{\Rs} &\text{if $r\leq \Rs \cos\epsilon$,}\\
\cos \epsilon &\text{if $r>\Rs \cos\epsilon$,}
\end{dcases} \subtag{c}
\end{equation}
for wedge angle $2\epsilon$. (Note that the wedge region is introduced to avoid the possibility that $r(\xi)$ becomes non-monotonic once the membrane inclination angle becomes close to $\pi/2$.) In the out-of-contact region (\ie~$\Rin\leq r\leq\Rout$) we impose the (integrated) vertical force balance
\begin{equation}\label{eq:nl_outplaneforce}
rT_\xi \sin\alpha = \frac{F}{2\pi}, \subtag{d}
\end{equation}
where $T_\xi$ and $T_\phi$ are the thickness-averaged, in-plane, radial and azimuthal stresses, which must satisfy the in-plane force balance
\begin{equation}
\frac{\de\,}{\de r} \left[rT_\xi\right] = T_\phi. \subtag{e}
\end{equation}
\end{subequations}
Note that the out-of-plane stress is forced to be zero for thin sheets ($t\ll 1$) [\citealt{Green1960}].

In modelling finite deformations of a thin membrane, one must distinguish between the reference and deformed configurations. Here, we take the reference configuration to be the planar sheet, subjected to an isotropic tension $\Tpre$; this configuration is parametrized by $(\rr, 0)$ for $0\leq\rr\leq \Rout$ (where  clamping is imposed at $\rr=r=\Rout$). The  variables describing the deformed configuration are expressed as functions of $\rr$ and hence we define the principal stretches as
\begin{subequations}\label{eq:nl_stretches}
\begin{equation}
\lambda_\xi \coloneqq \frac{\de \xi}{\de \rr}, \qquad \lambda_\phi \coloneqq  \frac{r}{\rr}, \qquad \lambda_z \coloneqq  \frac{t}{t_0}. \subtag{a--c}
\end{equation}
\end{subequations}
Here $\lambda_\xi$ is the longitudinal stretch (along a cross-sectional curve in the $r$--$z$ plane), $\lambda_\phi$ is  latitudinal stretch (along the direction normal to $r$--$z$ plane), and $\lambda_z$ is the out-of-plane stretch (a measure of membrane thickness $t$ compared to its reference value $t_0$). We also introduce the planar and out-of-plane pre-stretches $\Lambda:=\Lambda_\xi=\Lambda_\phi$ and $\Lambda_z$ which measure the initial isotropic stretching of the sheet; we will relate these pre-stretches to the pre-tension $\Tpre$ shortly.  

 System \eqref{eq:nl_reqs} is closed  by imposing a constitutive relation that links the  stresses with the principal stretches \eqref{eq:nl_stretches}. In particular, by assuming a hyperelastic isotropic medium,  we can introduce a constitutive strain energy density (per unit-volume) function $W(\lambda_1, \lambda_2, \lambda_3)$ [see \eg~\citealt{Holzapfel2002}] which directly links the stretches and stresses.

Using this formalism, we first calculate the pre-stretch $\Lambda$ by solving  the initial stress state,
\begin{subequations}\label{eq:prestretch}
\begin{equation}
\frac{\hat{t}_0}{\Lambda}\frac{\partial W}{\partial \lambda_1}(\Lambda,\Lambda, \Lambda_z) \equiv \frac{\hat{t}_0}{\Lambda}\frac{\partial W}{\partial \lambda_2}(\Lambda,\Lambda, \Lambda_z)=\Tpre,
\end{equation}
with $\hat{t}_0$ being the undeformed sheet thickness and $\Lambda_z$ chosen to satisfy the zero out-of-plane stress condition
\begin{equation}
\frac{\partial W}{\partial \lambda_3}(\Lambda,\Lambda,\Lambda_z) = 0.
\end{equation}
\end{subequations}
The stresses in the deformed configuration are then computed using
\begin{subequations}\label{eq:stress_stretch}
\begin{align}
T_\xi &= \frac{\hat{t}_0}{\Lambda\lambda_\phi}\frac{\partial W}{\partial \lambda_1}(\Lambda\lambda_\xi,\Lambda\lambda_\phi,\Lambda_Z\lambda_z)  \quad
\text{and} \quad T_\phi =\frac{\hat{t}_0}{\Lambda\lambda_\xi}\frac{\partial W}{\partial \lambda_2}(\Lambda\lambda_\xi,\Lambda\lambda_\phi,\Lambda_Z\lambda_z), \subtag{a,b}
\end{align}
where $\lambda_z$ satisfies the zero out-of-plane stress condition
\begin{equation}
\frac{\partial W}{\partial \lambda_3}(\Lambda\lambda_\xi,\Lambda\lambda_\phi,\Lambda_Z\lambda_z) = 0. \subtag{c}
\end{equation}
\end{subequations}

Together, \eqref{eq:nl_reqs}--\eqref{eq:stress_stretch} form a system of three ordinary differential equations for the three unknowns ---  the in-plane stretches $\lambda_\xi(r)$ and $\lambda_\phi(r)$ and the vertical displacement $z(r)$:
\begin{subequations}\label{eq:nl_odesystem}
\begin{align}
\frac{\de \lambda_\xi}{\de r} &=\frac{(T_\phi-T_\xi)\lambda_\xi\cos\alpha - \lambda_\phi T_{\xi,\,\phi} (\lambda_\xi\cos\alpha-\lambda_\phi)}{r\lambda_\xi T_{\xi,\,\xi}\cos\alpha},\\
\frac{\de \lambda_\phi}{\de r} &=\left(1-\frac{\lambda_\phi}{\lambda_\xi\cos\alpha}\right)\frac{\lambda_\phi}{r},\\
\frac{\de z}{\de r} &= \tan\alpha,
\end{align}
\end{subequations}
where $\alpha(r)$  is given by \eqref{eq:sphere_wedge} for $0\leq r\leq \Rin$ and \eqref{eq:nl_outplaneforce} for $\Rin\leq r\leq \Rout$; and $T_\xi(\lambda_\xi,\lambda_\phi)$, $T_\phi(\lambda_\xi,\lambda_\phi)$, and $T_{\xi,\,j}(\lambda_\xi,\lambda_\phi) \coloneqq \de T_\xi/\de\lambda_j$ are given by \eqref{eq:prestretch} and \eqref{eq:stress_stretch}.  To proceed further requires a particular choice of strain energy function $W$, and so we turn to discuss this now.

\subsection{Choice of strain energy density function}

The choice of strain energy density function $W(\lambda_1,\lambda_2,\lambda_3)$ is informed by the  material of interest. In this paper, we  present results for two hyperelastic models to show the influence of this choice. In particular, we present results for a neo-Hookean material (the natural extension of the Hookean response that is implicit in the FvK equations) and for a Gent hyperelastic material (which is a model developed for polymeric materials with finite extensibility, but is chosen here as a qualitative way to account for the finite bond-lengths in graphene). The Gent model contains a parameter $b$ that captures the finite chain length and recovers the neo-Hookean strain energy function as $b\to0$ (corresponding to infinite chain extensibility). The formulation of the compressible strain energy density functions is a lengthy process --- the details are presented in \ref{app:strain_energy}. The final form of the Gent energy density function used is
\begin{equation}\label{eq:gent_strainenergy}
W = \frac{E}{4(1+\nu)}\bigg[-\frac{1}{b}\log\big[1-b(I_1-3)\big] +(\beta-b)(I_3-1)-(1+\beta-b)\log I_3\bigg],
\end{equation}
where $\beta \coloneqq \nu/(1-2\nu)$ is some known constant,  $\nu\neq1/2$ is the Poisson ratio, $I_i$ are the usual tensor invariants (defined in \ref{app:strain_energy}), and $b$ is an empirical parameter based on the finite extensibility of the material, defined such that $1/b\equiv\max\{I_1-3\}$.

\subsection{Boundary conditions}

Having split the domain  into contacting  ($0\leq r\leq\Rin$) and non-contacting regions ($\Rin\leq r\leq\Rout$), the problem is a multi-point boundary value problem and so we require boundary conditions at three positions $r=0$, $\Rin$, and $\Rout$.

\emph{At the outer edge, $r=\Rout$}: the sheet is perfectly clamped,
\begin{subequations}\label{eq:nl_bcs}
\begin{equation}
\lambda_\phi(\Rout)= 1 \quad \text{and} \quad  z(\Rout) = 0. \subtag{a,b} 
\end{equation} 

\emph{At the origin, $r=0$}: we require an isotropic stretch (due to the symmetry of the problem),
\begin{equation}
\lambda_\phi(0)= \lambda_\xi(0)\quad  \text{and} \quad  z(0) = -\delta. \subtag{c,d}
\end{equation}
\end{subequations}

\emph{At the interface between the contacting and non-contacting regions, $r=\Rin$}: a local force balance reveals that we require continuity in radial stress $T_\xi(\lambda_\xi,\lambda_\phi)$ and membrane slope $\alpha$. Coupling this with the physical requirement of continuity of deformed variables $z$ and  $r=\rr\lambda_\phi$, we require continuity in all our variables  $\alpha$, $z$, $\lambda_\xi$, and $\lambda_\phi$. [This would not have been the case if there was a corner in the imposed indenter geometry (\eg~for a cylindrical punch).] Note that continuity in $\alpha$ gives an extra equation for the unknown point of contact $\Rin(F)$.

\subsection{Non-dimensionalization}
To non-dimensionalize the problem we use the same choice of dimensionless radius and vertical coordinate ($\rd$ and $Z$)  as the F\"oppl--von K\'arm\'an formulation \myeqref{eq:dimless_var}{a,c}, and define also
\begin{subequations}\label{eq:dimless_var_nl}
\begin{equation}
\rd \coloneqq \frac{\rr}{\Rout},  \qquad \hat{T}_\xi \coloneqq \frac{T_\xi}{\Tpre}, \qquad \hat{T}_\phi \coloneqq \frac{T_\phi}{\Tpre}, \subtag{a--c} 
\end{equation}
and
\begin{equation}
\hat{W}(\lambda_\xi,\lambda_\phi,\lambda_z) \coloneqq \frac{W(\lambda_\xi,\lambda_\phi,\lambda_z)}{E}, \subtag{d}
\end{equation}
\end{subequations}
were we use $\Y \coloneqq \hat{t}_0 E$ to be the two-dimensional Young's Modulus of the undeformed  sheet for small strains --- this is equivalent to the FvK choice under small strains. 

Substitution of \myeqref{eq:dimless_var}{a,c} and \eqref{eq:dimless_var_nl} into equations \myeqref{eq:nl_reqs}{c,d},  \eqref{eq:nl_odesystem}, and \eqref{eq:nl_bcs} gives a dimensionless system for $\lambda_\xi(\rd)$, $\lambda_\phi(\rd)$, $Z(\rd)$, and  $\alpha(\rd)$, with Cauchy stresses  given by
\begin{subequations}\label{eq:dimensionless_stress}
\begin{align}
\hat{T}_\xi &= \frac{\myT^{-1}}{\Lambda\lambda_\phi}\frac{\partial \hat{W}}{\partial \lambda_1}(\Lambda\lambda_\xi,\Lambda\lambda_\phi,\Lambda_Z\lambda_z),\\ 
\hat{T}_\phi &=\frac{\myT^{-1}}{\Lambda\lambda_\xi}\frac{\partial \hat{W}}{\partial \lambda_2}(\Lambda\lambda_\xi,\Lambda\lambda_\phi,\Lambda_Z\lambda_z), \\
0&=\frac{\partial \hat{W}}{\partial \lambda_3}(\Lambda\lambda_\xi,\Lambda\lambda_\phi,\Lambda_Z\lambda_z),
\end{align}
where the principal stretches $\Lambda$ and $\Lambda_z$ solve
\begin{equation}
 \frac{1}{\Lambda}\frac{\partial \hat{W}}{\partial \lambda_1}(\Lambda,\Lambda, \Lambda_z) \equiv \frac{1}{\Lambda}\frac{\partial \hat{W}}{\partial \lambda_2}(\Lambda,\Lambda, \Lambda_z)=\myT \quad \text{and} \quad
\frac{\partial \hat{W}}{\partial \lambda_3}(\Lambda,\Lambda,\Lambda_z)=0. \subtag{d,e}
\end{equation}
\end{subequations}

For a given strain energy density function $\hat{W}(\lambda_1,\lambda_2,\lambda_3)$ [we use \eqref{eq:gent_strainenergy}] the above system can be solved by a standard numerical integrator (in our work we use \texttt{bvp4c} in \texttt{\textsc{Matlab}}) for given parameters $\nu$, $\myF$, $\myT$, and $\myRs$, with unknown $\myR$. The associated indentation depth, $\myd(\myF)$, can then be calculated from \myeqref{eq:nl_bcs}{d}. In practice, however, it is move convenient to impose $\myR$ instead of $\myF$ and use the first derivative of \eqref{eq:nl_outplaneforce} to form a differential equation for the out-of-contact $\alpha$:
\begin{equation}
\frac{\de \alpha}{\de \rd}= - \frac{T_\phi\tan\alpha}{\rd T_\xi},
\end{equation}
in $\myR\leq\rd\leq1$, with a continuity boundary condition at $\rho=\myR$; $\myF$ can then be extracted post computation, along with $\myd$. By doing so, one avoids the issue of unknown domain size.

\subsection{Results}

\begin{figure}[ht!]
\centering
\includegraphics[width=0.9\textwidth]{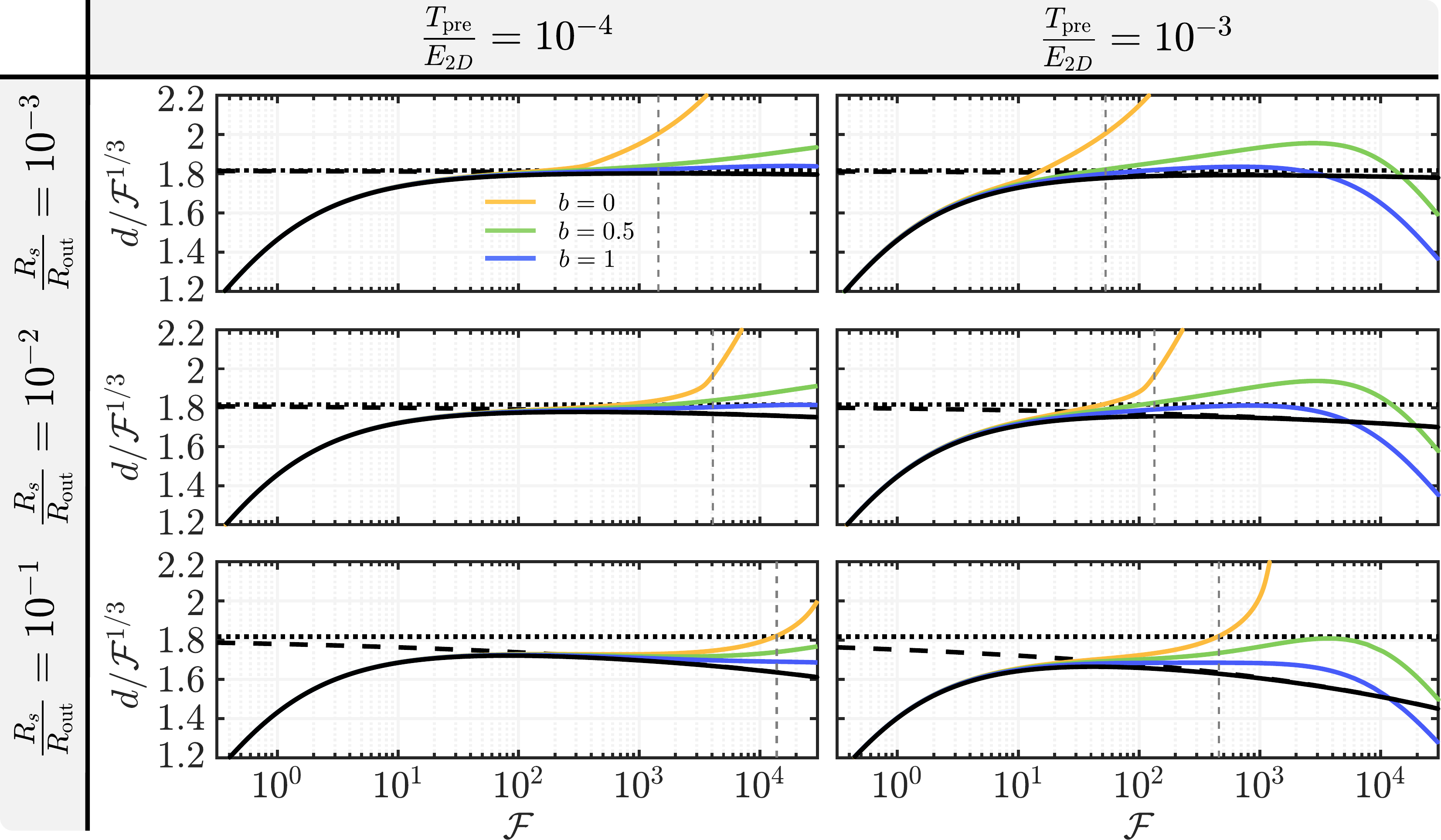}
\caption{Table of plots showing how the onset of nonlinear elasticity depends on the spherical cap curvature (rows) and the applied pre-tension (columns) of the membrane ($\nu =1/3$, $\myB=0$) --- here we have used an indenter wedge angle of $2\epsilon=\pi/9$ (matching the value in \eg~\citealt{LopezPolin2017}). Numerical solutions are shown as solid curves: F\"oppl--von K\'arm\'an (linear elastic) in black, Gent ($b=1$) in blue, Gent ($b=0.5$) in green, and neo-Hookean ($b=0$) in yellow. The asymptotic results from the FvK analysis for a point  and spherical indenter,  \eqref{eq:cubic_sph} with $\myRs=0$ and $\myRs=(\Rs/\Rout)(\Tpre/\Y)^{1/2}$, are shown as dotted and dashed lines respectively. The vertical dashed lines show where the relative difference between the Neo-Hookean and FvK models reaches 10\%.
\label{fig:stretch_NL_sph}}
\end{figure}

The aim of introducing a nonlinear elastic model was to investigate when the effects of material (as opposed to geometric) nonlinearity are observed in the key force--displacement curve. Although the quantitative results are highly dependent on the chosen constitutive strain energy density function   \eqref{eq:gent_strainenergy}, the transition  from the linearized-material asymptotics of \S\ref{sec:sphere}  occurs at a similar indentation force, independent of the choice of strain energy function. We are therefore able to investigate numerically the effect of varying the sphere radius $\myRs=(\Rs/\Rout)\myT^{1/2} $ and pre-tension $\myT$. In  Fig.~\ref{fig:stretch_NL_sph} we present a table of force--displacement curves obtained as $\myT$ and $\myRs$ vary; observe that the dimensionless force at which the results deviate from the F\"oppl--von K\'arm\'an solutions increases with $\Rs/\Rout$ and decreases with $\myT=\Tpre/\Y$ --- reminiscent of the prediction from the energetic analyses,  $\myF\sim\myRs/\myT^{2}$, presented in \S\ref{sec:scalings}.  It is also interesting to note that, close to the place at which the non-Hookean results deviate from the FvK results, the effect of material nonlinearity is to soften the material response.  This is in contrast to the geometrical nonlinearities discussed in \S\ref{sec:sphere}, which acted to increase the instantaneous estimate of $\Y$, \emph{i.e.}~to stiffen the indentation response. We discuss these results, and their significance for indentation probes of the elastic constants of thin materials, now.

\section{Discussion: Application to fitting protocols \label{sec:discussion}}

The controlled indentation of thin sheets is a common, but delicate, experimental technique used to extract mechanical properties of thin, approximately two-dimensional, materials. Our detailed analysis of the cylindrical and spherical indentation has led to a number of asymptotic results that highlight the complexity of this problem (these results are summarized in \S\ref{sec:summary}). Consequently, there are a number of potential pitfalls in this fitting procedure that must be appreciated if they are to be avoided and a reliable measurement of the quantity of interest to be obtained. Below, we  discuss these pitfalls in the context of measurements of three properties: the sheet pre-tension $\Tpre$, the two-dimensional Young's modulus $\Y$, and  the non-Hookean material behaviour (\ie~the behaviour not governed by linear elasticity). We shall concentrate on spherical-capped indenters, since these are among the most commonly used in practice.

\subsection{Measuring sheet pre-tension}

Our results showed the existence of a small-indentation regime in which the pre-tension dominates and $F\propto \Tpre \delta$ (Fig.~\ref{fig:regime_cylsph}). Based on our analysis of this region, there are two potential pitfalls that may cause errors when attempting to infer $\Tpre$ from experimental measurements of $F$:

\begin{itemize}

\item  \emph{Bending stiffness.} Although  the bending stiffness is small at a macroscopic level, \ie~$B\ll\Tpre\Rout^2$, it is not necessarily negligible in the early stages of indentation with a spherical-capped indenter (see Fig.~\myref{fig:regime_cylsph}{b}):  at  small indentations, the  spherical cap is barely wrapped by the sheet, the  contact radius is small and it is the bending stiffness of the sheet that dominates through an effective, bending-induced, radius $\Rin^{\mathrm{eff}}\sim(B/\Tpre)^{1/2}$, instead of the contact radius.  Avoiding this requires that the applied force be sufficiently large; in particular, that $F\gg B/\Rs$ for small bending stiffnesses ($B^{1/2}\ll \Rs \Tpre/\Y^{1/2}$) and $F \gg B^{1/2}\Tpre/\Y^{1/2}$ for moderate bending stiffnesses ($\Rs \Tpre/\Y^{1/2}\ll B^{1/2} \ll \Rout\Tpre^{1/2}$).

\item  \emph{Sheet stretching.} If the  indentation depths used are not sufficiently small, the pre-tension might be insignificant compared to the tension associated with indentation-induced stretching. This would cause a non--linear response (\ie~the sheet  transitions from the linear $\myF\sim\myd$ towards the cubic $\myF\sim\myd^3$ behaviour). Avoiding this  requires that the indentation force is not too large, in particular that $F\ll \Rout\Tpre^{3/2}/\Y^{1/2}$.

\end{itemize}

Combining these requirements, we find that indentation tests aimed at measuring the pre-tension in a sheet should focus on indentation forces $\FT$ such that
\begin{equation}\label{eq:preten_trap}
\min\left\{\frac{B}{\Rs}, \, \frac{B ^{1/2}\Tpre}{\Y^{1/2}}\right\}\ll\FT\ll \frac{\Rout\Tpre^{3/2}}{\Y^{1/2}}.
\end{equation} 

\subsection{Measuring 2D Young's modulus}

The measurement of the  two-dimensional Young's modulus of a sheet requires experiments to be performed in the stretching dominated regime,  where $F\propto \Y \delta^3/\Rout^2$ (Fig.~\ref{fig:regime_cylsph}). Based on our analysis of this regime, there are two immediate potential pitfalls that may cause fitting errors:

\begin{itemize}

\item \emph{Sheet pre-tension.} If the sheet is not sufficiently indented, the effect of the pre-tension may still be significant, leading to a non-cubic response (\ie~the sheet is still transitioning between the linear $\myF\sim \myd$ and cubic $\myF\sim \myd^3$ behaviours, see Fig.~\ref{fig:stretch_NL_sph}). To avoid this, sufficiently large forces should be applied; in particular $F \gg \Rout\Tpre^{3/2}/\Y^{1/2}$.

\item \emph{Mechanical nonlinearities.} If the sheet is indented too much, the stress in the sheet exceeds that for which the linear elastic (Hookean) constitutive response is valid, and nonlinearities become important. In this case, the expected cubic response may not be observed (see the solution divergence in  Fig.~\ref{fig:stretch_NL_sph}) causing errors in the fitted value of $\Y$. To ensure this possibility is avoided requires $F \ll \Y\Rs$.

\end{itemize}

Combining these requirements, we find that indentation tests aimed at measuring the stretching modulus of a sheet should focus on indentation forces $\FY$ such that
\begin{equation}\label{eq:YM_trap}
\frac{\Rout\Tpre^{3/2}}{\Y^{1/2}}\ll \FY \ll \Y\Rs.
\end{equation}

A third potential pitfall is less obvious and independent of the indentation force applied:
\begin{itemize}
\item \emph{Geometrical nonlinearities.} When fitting experimental data to obtain a value for the 2D Young's modulus, it is common practice to use the \cite{Schwerin1929} point indenter solution, which in our notation reads $\myF=Q(\nu)\myd^3$ for $Q(\nu)\coloneqq q_p[\nu]^{-3}$. For example, a point-wise estimate $\Y=F\Rout^2/\bigl[2\pi Q(\nu)\delta^3\bigr]$ is often used (\eg~in \citealt{LopezPolin2017}). However, the analysis presented here shows that the indenter geometry is important and may lead to large errors in the fitted value of the 2D Young's modulus if not accounted for. In particular, for  spherical indenters the point-wise estimate of $\Y$ may not converge as the force increases ---  the constant value (as predicted by \citeauthor{Schwerin1929}) is never observed (see Figs~\ref{fig:fvkcubic_cyl_sph}~\&~\ref{fig:stretch_NL_sph}). To avoid a significant effect of the spherical geometry  requires $\Rs/\Rout \ll 1$; however, we caution that decreasing $\Rs$ or increasing $\Rout$ may have the undesired effect of decreasing the desired range of indentation forces in \eqref{eq:YM_trap}, risking other fitting errors. In practice, $\Rs/\Rout\lesssim 10^{-2}$ appears to be sufficient (see Fig.~\ref{fig:stretch_NL_sph}); in many experimental setups $\Rs/\Rout\sim0.1$ \cite[see][for example]{LopezPolin2017} suggesting that  indenter geometry may play a role in interpreting previous experimental results. 

\end{itemize}

\subsection{Measuring non-Hookean material behaviour\label{sec:MeasureNonHookean}}

If interested in examining the non-Hookean material behaviour of the sheet (\ie~the behaviour beyond the Hookean linear stress-strain relation), there are two potential pitfalls to avoid:

\begin{itemize}

\item \emph{Remaining in the linear elastic regime.} The main concern when measuring non-Hookean behaviour is whether the strains induced by indentation are large enough to be controlled by a nonlinear constitutive law. For a spherical-capped indenter, this requires
\begin{equation}\label{eq:nonlin_trap}
\Y \Rs \ll F.
\end{equation}

\item \emph{Geometrical nonlinearities.} A less obvious trap is that deviating from \citeauthor{Schwerin1929}'s solution [$\myF=Q(\nu)\myd^3$] might be interpreted as nonlinear mechanical effects, but actually result from nonlinear geometry. For example, for sufficiently large spherical indenters, even the \fvk response will never achieve the plateau expected from the \citeauthor{Schwerin1929} result (see Figs \ref{fig:fvkcubic_cyl_sph}~\&~\ref{fig:stretch_NL_sph}). This failure to reach the Schwerin regime might be interpreted as a material nonlinearity, rather than a universal geometric property that is predicted by Hookean elasticity. To avoid this possibility requires geometric effects to be negligible throughout the linear elastic regime, that is $\Rs \ll \Rout$. However, as we now discuss, our results suggest that the effect of geometrical and material nonlinearities on the cubic compliance $d/\myF^{1/3}$ are qualitatively different.

\end{itemize}

\subsection{Distinguishing geometrical and material nonlinearities}

Together,  \eqref{eq:YM_trap} and \eqref{eq:nonlin_trap} provide limits on the validity of fitting force--indentation curves to determine the stretching modulus $\Y$ of a two-dimensional material, and to distinguish between geometrical and material nonlinearities in this procedure. (Of course, their validity can only be checked with some initial parameter estimate or \emph{a posteriori}.) It should be noted that these are only based on order arguments and are not concrete cut off points. For example, the numerical results presented in Figs~\ref{fig:fvkcubic_cyl_sph}~\&~\ref{fig:stretch_NL_sph} suggest that to observe the asymptotic results for $\myF\gg1$ in practice requires $\myF\gtrsim 100$; hence a useful guide for satisfying the conditions \eqref{eq:YM_trap} in practice is
\begin{equation}\label{eq:YM_trap_2}
200\pi\times\frac{\Rout\Tpre^{3/2}}{\Y^{1/2}}\lesssim\FY\ll\Y\Rs.
\end{equation}
Comparison of the relevant experimental parameters collected in Table~\ref{tab:experiments} with \eqref{eq:YM_trap_2} suggests that the maximum indentation force applied experimentally, $F_{\max}$, does not always reach the large multiple of $\Tpre^{3/2}\Rout/\Y^{1/2}$ required for accurate measures of $\Y$ from asymptotic results. 

Hence,  when fitting, one should always ensure the desired asymptotic behaviour is observed. To fit the stretching stiffness $\Y$, this might be most simply done by ensuring that there is indeed a cubic plateau, $F/d^3\sim \text{constant}$. However, our results show that a  true plateau is only obtained with a cylindrical indenter, while most experiments use a shape that is closer to a spherical cap. Fortunately, qualitatively similar asymptotic results may be derived from the \fvk equations for a spherical-capped indenter; in this regard \myeqref{eq:cubic_sph_full}{a} might be expected to be especially helpful and so we note that it may be rewritten in dimensional terms as
\beq
\frac{\delta}{\Rout}\sim\frac{q_p[\nu]}{(2\pi)^{1/3}}\left(\frac{F}{\Y\Rout}\right)^{1/3}-\frac{2(\sqrt{2}-1)}{\pi^{1/2}}\left(\frac{\Rs}{\Rout}\right)^{1/2}\left(\frac{F}{\Y\Rout}\right)^{1/2}.
\label{eqn:SphereDim}
\eeq 
Note that \eqref{eqn:SphereDim} reduces to the classical \citeauthor{Schwerin1929} result as $\Rs/\Rout\to0$ but shows that the perturbation caused by spherical geometry is \emph{not} of the prefactor \cite[as assumed by][for example]{Begley2004} but rather is additive. Moreover,  \eqref{eqn:SphereDim} shows that the cubic compliance $\delta/F^{1/3}$ is decreased by the effect of the indenter's radius of curvature $\Rs$: the geometry of the indenter means that the apparent stiffness of the suspended solid is \emph{increased} compared to a point indenter, and increases further with increasing load. As a rule of thumb, our results suggest that material nonlinearities tend to soften the response initially (at least for the strain energy functionals considered here) while geometrical nonlinearities tend to stiffen the response.

\section{Summary of results and conclusions}\label{sec:summary}

\subsection{Summary}

We have presented a series of asymptotic solutions that may be used in combination with experimental force--indentation data to fit mechanical properties of thin materials, subject to an appreciation of the pitfalls described in \S\ref{sec:discussion}.  These results apply in various asymptotic regimes determined by the relative sizes of the dimensionless force $\myF$, bending stiffness $\myB$, indenter radius $\myR$ (or $\myRs$), and pre-tension $\myT$. Here, we summarize these  results for a cylindrical and spherical-capped indenter; these are most easily expressed in terms of the vertical indentation depth $d$ achieved for a fixed force $\myF$.

\paragraph{Cylindrical indenter} For indentation by a cylindrical indenter,  the asymptotic response is dependent on the relative size of the bending stiffness $\myB$ and indenter area $\myR^2$. We have that
\begin{subequations}\label{eq:dF_overall_cyl}
\begin{align}
&\text{if $\myB^{1/2}\ll \myR<1$:}&   &\myd\sim
\begin{dcases}
\myF\log\frac{1}{\myR} & \text{for }0\leq\myF\ll \myR,\\
\myF\log\frac{4}{\myF} & \text{for } \myR\ll\myF\ll 1,\\
\myF^{1/3} q_c[\nu,\myR]& \text{for }1\ll\myF,
\end{dcases}
\intertext{whilst}
&\text{ if $\myR\ll \myB^{1/2}\ll1$:}& &\myd\sim
\begin{dcases}
\myF \log\frac{e^\gamma}{2\myB^{1/2}} & \text{for }0\leq\myF\ll \myB^{1/2},\\
\myF\log \frac{4}{\myF} & \text{for } \myB^{1/2}\ll\myF \ll 1,\\
 \myF^{1/3} q_c[\nu,\myR] & \text{for }1\ll\myF.
\end{dcases}
\end{align}
\end{subequations}
Here, $\gamma\approx 0.577$ is the Euler--Mascheroni constant \cite[][]{Abramowitz1964} and $q_c[\nu,\myR]\sim q_p[\nu]-(2\myR)^{2/3}$ is defined in \eqref{eq:cubic_cyl}. These solutions are valid provided $\myT\ll 1$ and $\myF\ll \myR\myT^{-3/2}$; if either of these conditions fail then the material instead behaves according to a nonlinear constitutive law (non-Hookean behaviour). The regions of the regime diagram Fig.~\myref{fig:regime_cylsph}{a} are delineated by the expressions in \eqref{eq:dF_overall_cyl}.

\paragraph{Spherical indenter} For indentation by a spherical-capped indenter, we find the asymptotic response:
\begin{subequations}\label{eq:dF_overall_sph}
\begin{align}
\myd &\sim \begin{dcases}
\myF \log\frac{e^\gamma}{2\myB^{1/2}} & \text{for }0\leq \myF\ll\min\{\myB/\myRs,\myB^{1/2}\},\\
\myF\log\sqrt{\frac{e}{\myRs\myF}} & \text{for }\myB/\myRs\ll\myF\ll \myRs,\\
\myF\log\frac{4}{\myF}  & \text{for }\max\{\myRs,\myB^{1/2}\}\ll\myF\ll 1,\\
\myF^{1/3} q_s[\nu, \myRs^3\myF]   & \text{for } 1\ll\myF,
\end{dcases}
\intertext{with the (\emph{a priori} unknown) contact radius given by}
\myR^2 &\sim
\begin{dcases}
4\myB e^{-2\gamma}\exp[-4\myB/\myRs\myF] & \text{for }0\leq \myF\ll\min\{\myB/\myRs,\myB^{1/2}\},\\
\myRs\myF & \text{for }\myB/\myRs\ll\myF\ll \myRs,\\
4(\sqrt{2}-1)\myRs^{3/2}\myF^{1/2} &  \text{for } \max\{\myRs,\myB^{1/2}\}\ll\myF.
\end{dcases}
\end{align}
\end{subequations}
Here,  $q_s[\nu,\myRs^3\myF]\sim q_p[\nu]-2(2-\sqrt{2})(\myRs^3\myF)^{1/6}$ is defined in \eqref{eq:cubic_sph_full} and we note that the dimensional version of \myeqref{eq:dF_overall_sph}{a} for $\myF\gg1$ is given in \eqref{eqn:SphereDim}. These solutions are valid provided $\myT\ll 1$ and $\myF\ll\myRs\myT^{-2}$; if either of these conditions fail then the material instead behaves according to a nonlinear constitutive law (non-Hookean behaviour). The regions of the regime diagram Fig.~\myref{fig:regime_cylsph}{b} are delineated by the expressions in \eqref{eq:dF_overall_sph}.  
 
\subsection{Conclusions}
 
 Altogether, our work  provides a comprehensive description of the Hookean response of a sheet subject to localized indentation accounting for geometrical nonlinearities, while additionally providing information about when a non-Hookean response can be expected.  It is common in experiments to use  dimensional versions of similar asymptotic solutions to extract information about the material of interest (\eg~its 2D Young's modulus, $\Y$, or pre-tension, $\Tpre$). The asymptotic results presented in \eqref{eq:dF_overall_cyl} and \eqref{eq:dF_overall_sph} show the number of different regimes that exist and hence the difficulty of choosing the appropriate asymptotic result. Nevertheless, understanding the appropriate regime for each of these results is important, since incorrect choices may lead to large errors in the fitted values obtained.  For instance,  if the indenter was assumed to be point-like (so that $q_c[\nu,\myR] \approx q_p[\nu]$ or $q_s[\nu,\myRs^3\myF] \approx q_p[\nu]$), we would obtain $\Oh(\myR^{2/3})$ or  $\Oh(\myRs^{1/2}\myF_{\max}^{1/6})$ errors in the fitted Young's modulus --- which can be significant in such a sensitive process. We therefore emphasize the importance of using the correct response when fitting parameters and suggest the use of \eqref{eqn:SphereDim} to account for indenters with a hemi-spherical tip.
  
Finally, we note that the asymptotic regimes considered here have been motivated by recent indentation experiments on ultra-thin materials, including few layer graphene. For such materials,  $B\ll \Tpre \Rout^2$, corresponding to $\myB\ll1$, and so the different regimes described in eqns \eqref{eq:dF_overall_cyl} and \eqref{eq:dF_overall_sph} are only valid when $\myB\ll1$. As the dimensionless bending stiffness becomes larger, $\myB=O(1)$, the balances that lead to these results are expected to change. In particular, we expect that for dimensionless forces $\myF\lesssim1$ and $\myB\sim1$ new results would be required (indicated by the greyed-out regions in Fig.~\ref{fig:regime_cylsph}).

\section*{Acknowledgments}
The research leading to these results has received funding from the European Research Council under the European Union's Horizon 2020 Programme/ERC Grant No.~637334 (D.V.), a Philip Leverhulme Prize (D.V.) and the EPSRC Grant No.~EP/M508111/1 (T.C.).  The numerical data that supports the plots within this paper are available to download from https://doi.org/10.5287/\\ bodleian:dmKYJKX1z. This work was performed in part at Aspen Center for Physics, which is supported by National Science Foundation grant PHY-1607611. We are grateful to Cristina G\'omez-Navarro for bringing the work of \cite{Jin2017} to our attention.

\appendix

\section{Asymptotic solutions for small indentation forces}
\label{app:small_forces}

In this Appendix, we extend the arguments of \S\ref{sec:results_B>0_cyl} and \S\ref{sec:results_B>0_sph} to derive the asymptotic solutions \eqref{eq:sol_linbencyl} and \eqref{eq:sph_linben_full}. We shall present the analysis for the cylindrical and spherical-capped indenters simultaneously.

We consider a small perturbation of the initial pre-tensed configuration: $\Psi(\rd)\sim \rd + \tilde{\Psi}(\rd)$ and $Z(\rd)\sim \tilde{Z}(\rd)$, anticipating that $\tilde{\Psi},\tilde{Z}\ll 1$. At leading order (in $\tilde{\Psi}$ and $\tilde{Z}$), the F\"oppl--von K\'arm\'an equations \eqref{eq:fvk_dless} take the form,
\begin{subequations}\begin{align}
\myB \rd\frac{\de\,}{\de \rd} \left[\frac{1}{\rd} \frac{\de\,}{\de \rd}\left(\rd\frac{\de \tilde{Z}}{\de \rd}\right)\right] &= \rd\frac{\de \tilde{Z}}{\de \rd} -\myF,\label{eq:perturb_eq_1}\\
\rd\frac{\de\,}{\de \rd} \left[\frac{1}{\rd} \frac{\de \,}{\de \rd}\left(\rd\tilde{\Psi}\right)\right] &= -\frac{1}{2}\left(\frac{\de \tilde{Z}}{\de \rd}\right)^2.\label{eq:perturb_eq_2}
\end{align}\end{subequations}
Most importantly \eqref{eq:perturb_eq_1} no longer couples the out-of-plane deflection with the in-plane stress; the third order differential equation for the sheet profile $Z(\rd)\sim\tilde{Z}(\rd)$  can immediately be integrated to give
\begin{equation}\label{eq:Z_perturb_sol}
Z\sim \tilde{Z} = c_1 I_0\left[\frac{\rd}{\myB^{1/2}}\right] +c_2 K_0 \left[\frac{\rd}{\myB^{1/2}}\right] + \myF \log \rd + c_3,
\end{equation}
with the constants $c_1$, $c_2$, and $c_3$ yet to be determined.  (The stress profile may be determined by substituting \eqref{eq:Z_perturb_sol}  into \eqref{eq:perturb_eq_2}; we  omit this here since it affects the force--indentation response only at higher order in $\myd$.)

The sheet profile \eqref{eq:Z_perturb_sol} is to be solved subject to the boundary conditions
\begin{subequations}\label{eq:B>0_BC}
\begin{equation}
Z(1)=0, \qquad Z'(1)=0, \subtag{a,b}
\end{equation}
\end{subequations}
with
\begin{subequations}\label{eq:B>0_BC_cyl}
\begin{equation}
Z(\myR)=-\myd,\qquad Z'(\myR)=0, \subtag{a,b}
\end{equation}
\end{subequations}
for a  cylindrical indenter ($\myR$ known), or
\begin{subequations}\label{eq:B>0_BC_sph}
\begin{equation}
Z(\myR)=-\myd + \frac{\myR^2}{2\myRs},\qquad Z'(\myR)=\frac{\myR}{\myRs}, \qquad Z''(\myR)=\frac{1}{\myRs},\subtag{a--c}
\end{equation}
\end{subequations}
for a  spherical indenter ($\myR$ unknown). Implementation of \eqref{eq:B>0_BC}--\eqref{eq:B>0_BC_sph} gives explicit expressions for the coefficients
\begin{subequations}
\begin{equation}
\frac{c_1}{\myF\myB^{1/2}} = \frac{ \hat{K}_1^{R}-\hat{K}_1/\myR}{\hat{I}_1^{R}\hat{K}_1-\hat{I}_1\hat{K}_1^{R}}, \qquad
\frac{c_2}{\myF\myB^{1/2}}  = \frac{ \hat{I}_1^{R}-\hat{I}_1/\myR}{\hat{I}_1^{R}\hat{K}_1-\hat{I}_1\hat{K}_1^{R}},\subtag{a,b}
\end{equation}
and 
\begin{equation}
c_3 = -c_1 \hat{I}_0 - c_2\hat{K}_0, \subtag{c}
\end{equation}
\end{subequations}
for a cylindrical indenter. For a spherical indenter, we have
\begin{subequations}
\begin{equation}
\frac{c_1}{\myF\myB^{1/2}}  = \frac{\hat{K}_1\myR/\myF\myRs +\hat{K}_1^{R}-\hat{K}_1/\myR}{\hat{I}_1^{R}\hat{K}_1-\hat{I}_1\hat{K}_1^{R}},\qquad
 \frac{c_2}{\myF\myB^{1/2}} = \frac{\hat{I}_1\myR/\myF\myRs  + \hat{I}_1^{R}-\hat{I}_1/\myR}{\hat{I}_1^{R}\hat{K}_1-\hat{I}_1\hat{K}_1^{R}}, \subtag{a,b}
\end{equation}
and 
\begin{equation}
c_3 = -c_1 \hat{I}_0 - c_2\hat{K}_0, \subtag{c}
\end{equation}
with $\myR$ implicitly given by
\begin{equation}
\frac{1}{\myRs}+  \frac{\myF}{\myR^2} = \frac{1}{2\myB}\left[c_1\left(\hat{I}_0^R+\hat{I}_2^R\right)+ c_2\left(\hat{K}_0^R+\hat{K}_2^R\right)\right]. \subtag{d}
\end{equation}
\end{subequations}
(In each case,  $\hat{I}_j$, $\hat{I}^R_j$, $\hat{K}_j$, and $\hat{K}^R_j$ are as defined in eq.~\eqref{eq:myIK}.)

The small force--displacement relations are then found by imposing the boundary conditions \myeqref{eq:B>0_BC_cyl}{a} and \myeqref{eq:B>0_BC_sph}{a}, to give:
\begin{equation}\label{eq:bessel_cylsol_full}
\frac{\myd}{\myF} =  \log\frac{1}{\myR}+\frac{\hat{K}_1\hat{I}^R_0+ \hat{I}_1\hat{K}^R_0 +\myR\left(\hat{K}_0\hat{I}^R_1 +\hat{I}_0\hat{K}^R_1\right)-2\myB^{1/2}}{\hat{I}^R_1 \hat{K}_1 - \hat{I}_1\hat{K}^R_1}\frac{\myB^{1/2}}{\myR},
\end{equation}
for a cylindrical indenter; and 
\begin{subequations}\label{eq:bessel_sphsol_full}
\begin{equation}
\begin{split}
\frac{\myd}{\myF} = \log\frac{1}{\myR}&+\frac{\hat{K}_1\hat{I}^R_0+ \hat{I}_1\hat{K}^R_0 +\myR\left(\hat{K}_0\hat{I}^R_1 +\hat{I}_0\hat{K}^R_1\right)-2\myB^{1/2}}{\hat{I}^R_1 \hat{K}_1 - \hat{I}_1\hat{K}^R_1}\frac{\myB^{1/2}}{\myR}\\
 &+\frac{\myB^{1/2}-\hat{I}_1\hat{K}^R_0-\hat{I}^R_0\hat{K}_1}{\hat{I}^R_1 \hat{K}_1 - \hat{I}_1\hat{K}^R_1}\frac{\myR\myB^{1/2}}{\myRs\myF}+ \frac{\myR^2}{2\myRs\myF},
 \end{split}
\end{equation}
with $\myR$ implicitly given by
\begin{equation}
\frac{\myF\myRs +\myB^{-1/2}\left(\myR^2-\myF\myRs\right)\left(\hat{I}^R_0 \hat{K}_1+\hat{K}^R_0 \hat{I}_1 \right)}{\hat{I}^R_1\hat{K}_1-\hat{I}_1\hat{K}^R_1}= 2\myR,
\end{equation}
\end{subequations}
for a spherical indenter.

In the limit of small bending stiffnesses which is relevant here, $\myB\ll 1$,  \eqref{eq:bessel_cylsol_full} and \eqref{eq:bessel_sphsol_full} can be simplified using  the asymptotic behaviour of modified Bessel functions to give
\begin{equation}
\frac{\myd}{\myF} =  \log\frac{1}{\myR}-\myB^{1/2}\frac{\hat{K}^R_0}{\myR\hat{K}^R_1}-\myB^{1/2}\frac{\hat{I}_0}{\hat{I}_1}+ \text{e.s.t.},
\end{equation}
for a cylindrical indenter. For a spherical indenter  
\begin{subequations}
\begin{equation}
\frac{\myd}{\myF} = \log\frac{1}{\myR}-\frac{2\myB}{\myRs\myF}+\frac{\myR^2}{2\myRs\myF}-\myB^{1/2}\frac{\hat{I}_0}{\hat{I}_1}+ \text{e.s.t.},
\end{equation}
where $\myR$ is given  implicitly by
\begin{equation}
\left(\myRs\myF-\myR^2\right)\frac{\hat{K}^R_0}{\myR\hat{K}^R_1}= 2\myB^{1/2}+ \text{e.s.t.}.
\end{equation}
\end{subequations}
Finally, considering the limits $\myB\ll\myR^2<1$ and  $\myR^2\ll\myB\ll1$ leads to the leading-order solutions \eqref{eq:sol_linbencyl} for a cylindrical  indenter and \eqref{eq:sph_linben_full} for a spherical-capped indenter.

\section{Asymptotic solutions for vanishing bending stiffnesses}
\label{app:small_bending}

In this Appendix, we expand the arguments of \S\ref{sec:results_B=0_cyl}, \S\ref{sec:results_B=0_sph1}, and \S\ref{sec:results_B=0_sph2} to derive the asymptotic solutions for large indentation depths, and negligible bending stiffnesses \eqref{eq:sol_linptcyl}, \eqref{eq:cubic_cyl}, \eqref{eq:sph_linpt_full}, and \eqref{eq:cubic_sph_full}. We follow the analysis of a point-indenter by \citet[App.~B]{Vella2017} and present the case of cylindrical and spherical-capped indenters simultaneously.

Following the discussion in \S\ref{sec:results_B=0_cyl}, we let $\eta = \rd^2$, $\Phi = \rd\Psi$ and set $\myB=0$ in \eqref{eq:fvk_dless}, to give:
\begin{subequations}\label{eq:B0_eq_etaphi}
\begin{align}
2\Phi\frac{\de Z}{\de \eta}&=\myF,\label{eq:B0_eq_etaphi_1}\\
\frac{\de^2\Phi}{\de \eta^2} &= - \frac{1}{2}\left(\frac{\de Z}{\de \eta}\right)^2 = -\frac{\myF^2}{8\Phi^2}. \label{eq:B0_eq_etaphi_2}
\end{align}\end{subequations}
Neglecting $\myB$ reduces the order of the system, and so we suppress the boundary conditions on the highest-order quantities \myeqref{eq:bc_cyl_dless}{b,e} and \myeqref{eq:bc_sph_dless}{c}. \eqref{eq:B0_eq_etaphi} is, therefore, to be solved subject to
\begin{subequations}\label{eq:bc_Phieta1}
\begin{equation}
 2\Phi'(1)- (1+\nu) \Phi(1) = 1-\nu, \qquad Z(1) = 0, \subtag{a,b}
\end{equation} 
\end{subequations}
with:
\begin{subequations}\label{eq:bc_Phieta2}
\begin{equation}
 Z(\myR^2) = -\myd,\qquad \myR\Phi'(\myR^2)-  \frac{\Phi(\myR^2)}{\myR}= 0, \subtag{a,b}
\end{equation}
\end{subequations}
for a cylindrical indenter; or
\begin{subequations}\label{eq:bc_Phieta3}
\begin{equation}
 Z(\myR^2) = -\myd +\frac{\myR^2}{2\myRs}, \qquad Z'(\myR^2) =\frac{1}{2\myRs},\qquad
 \myR\Phi'(\myR^2)-\frac{\Phi(\myR^2)}{\myR} = -\frac{\myR^3}{16\myRs^2}, \subtag{a--c}
\end{equation}
\end{subequations}
for a spherical indenter. 

Two integrations of \eqref{eq:B0_eq_etaphi_2} leads to 
\begin{equation}\label{eq:Phi_sol}
(A\Phi)^{1/2}(1+A\Phi)^{1/2} -\sinh^{-1}\sqrt{A\Phi}= \frac{\myF A^{3/2}}{2}\eta + B,
\end{equation} for integration constants $A$ and $B$, while  \eqref{eq:B0_eq_etaphi_1} leads to
\begin{equation}\label{eq:Z_sol}
Z= \frac{2}{\sqrt{A}} \sinh^{-1}\sqrt{A\Phi} - C,
\end{equation}
for some constant $C$.

Applying the boundary conditions \eqref{eq:bc_Phieta1}--\eqref{eq:bc_Phieta3}  leads to a system for the unknown constants $A$, $B$, and $C$. To simplify these equations, we define $\Phi_0:=A\Phi(\myR^2)$ and $\Phi_1:=A\Phi(1)$; the system then takes the form
\begin{subequations}\label{eq:ABC_system}
\begin{align}
\Phi_1^{1/2}(1+\Phi_1)^{1/2} -\sinh^{-1}\Phi_1^{1/2} &= \frac{\myF A^{3/2}}{2}  + B,\\
\Phi_0^{1/2}(1+\Phi_0)^{1/2} -\sinh^{-1}\Phi_0^{1/2}&= \frac{\myF A^{3/2}}{2}\myR^2 + B,\\
\myF A^{3/2}\left(1+\frac{1}{\Phi_1}\right)^{1/2}-(1+\nu)\Phi_1 &= (1-\nu)A ,\label{eq:ABC_system_3}\\
 \frac{2}{A^{1/2}}\sinh^{-1}\Phi_1^{1/2} &= C,
\end{align}
\end{subequations}
with:
\begin{subequations}\label{eq:ABC_system_cyl}
\begin{align}
\myF A^{3/2}\left(1+\frac{1}{\Phi_0}\right)^{1/2} &= \frac{2\Phi_0}{\myR^2},\\
 \frac{2}{A^{1/2}} \sinh^{-1}\Phi_0^{1/2} &= C -d,
\end{align}
\end{subequations}
for a cylindrical indenter; or
\begin{subequations}\label{eq:ABC_system_sph}
\begin{align}
\myF A^{3/2}\left(1+\frac{1}{\Phi_0}\right)^{1/2} &=\frac{2\Phi_0}{\myR^2} -\frac{\myR^2A}{8\myRs^2},\\
 \frac{2}{A^{1/2}} \sinh^{-1}\Phi_0^{1/2} &=  C -d+\frac{\myR^2}{2\myRs} ,\\
 \Phi_0 &= A\myF\myRs,
\end{align}
\end{subequations}
 for a spherical indenter. Eliminating $B$ and $C$ from system \eqref{eq:ABC_system} with \eqref{eq:ABC_system_cyl} or \eqref{eq:ABC_system_sph}, leaves a force--displacement equation given implicitly by parameters $\Phi_0$, $\Phi_1$, and $A$ (with $\myR$ as an  extra unknown in the spherical case); the resulting system for a cylindrical indenter is given in  \eqref{eq:B0_full_cyl}. (Note that the parametric solution for a spherical indenter, \eqref{eq:ABC_system} with \eqref{eq:ABC_system_sph}, has been previously found by \citealt{Jin2017}; however, the following asymptotic results were not presented by them.)
 
We now consider the two limits $\myF\to 0$ and $\myF\to\infty$  to derive explicit asymptotic force--displacement relations;  in particular, equation \eqref{eq:ABC_system_3} determines how $A$ and $\Phi_1$ relate to the applied  force,  $\myF$.

\subsection{Moderate indentation forces $(\myF\to0)$}

In the small indentation limit, $\myF\to0$, eq.~\eqref{eq:ABC_system_3} gives  $A\sim\Phi_1= 4/\myF^2 +\Oh(1)$, which leads to the parametric force--displacement relationship for a cylindrical indenter:
\begin{subequations}
\begin{equation}
\frac{\myd}{\myF} = \log \frac{4}{\myF}-\sinh^{-1}\Phi_0^{1/2} +\Oh(\myF^2),
\end{equation}
where $\Phi_0$ solves
\begin{equation}
\frac{\Phi_0^{3/2}}{\left(1+\Phi_0\right)^{1/2}}\sim\frac{4\myR^2}{\myF^2}.
\end{equation}
\end{subequations}
For a spherical indenter, we find
\begin{subequations}
\begin{equation}
\frac{\myd}{\myF} = \log \frac{4}{\myF}-\sinh^{-1}\Phi_0^{1/2}+\frac{\myR^2}{2\myRs\myF} +\Oh(\myF^2),
\end{equation}
where $\Phi_0$ and $\myR$ are given by
\begin{align}
\Phi_0 &\sim \frac{4\myRs}{\myF},\\
\frac{\myR^2}{2\myRs\myF}&\sim\Phi_0^{1/2}\left\{(2+\Phi_0)^{1/2}-(1+\Phi_0)^{1/2}\right\}.
\end{align}
\end{subequations}
 Taking the limits $\myF\ll\myR<1$ and $\myR\ll\myF\ll1$ leaves the leading-order solutions  \eqref{eq:sol_linptcyl} for a cylindrical indenter and \eqref{eq:sph_linpt_full} for a spherical-capped indenter.

\subsection{Large indentation forces $(\myF\to \infty)$}
In the large indentation limit, $\myF\to\infty$, eq.~\eqref{eq:ABC_system_3} gives  $A=\Oh(\myF^{-2/3})$ and $\Phi_1=\Oh(1)$;  inserting these into the relevant system and rearranging leads to the parametric force--displacement relation for a cylindrical indenter:
\begin{subequations}\label{eq:cubic_cyl_full_app}
\begin{equation}
\frac{\myd}{\myF^{1/3}}  = \frac{2}{(1+\nu)^{1/3}}\frac{\sinh^{-1}\Phi_1^{1/2}-\sinh^{-1}\Phi_0^{1/2}}{\Phi_1^{1/2}\left(1+\Phi_1\right)^{-1/6}}+\Oh(\myF^{-2/3}),
\end{equation}
where $\Phi_1$ and $\Phi_0$ are given by
\begin{align}
 \frac{1+\nu}{2}\frac{\Phi_1^{3/2}}{(1+\Phi_1)^{1/2}}&\sim  \frac{\left[\Phi^{1/2}(1+\Phi)^{1/2}-\sinh^{-1}\Phi^{1/2} \right]_{\Phi_0}^{\Phi_1}}{1-\myR^2},\\
 \frac{1+\nu}{2}\frac{\Phi_1^{3/2}}{\left(1+\Phi_1\right)^{1/2}} &\sim \myR^{-2}\frac{\Phi_0^{3/2}}{\left(1+\Phi_0\right)^{1/2}}.
\end{align}
\end{subequations}
For a spherical indenter, we find
\begin{subequations}\label{eq:cubic_sph_full_app}
\begin{equation}
\frac{\myd}{\myF^{1/3}}  = \frac{2}{(1+\nu)^{1/3}}\frac{\sinh^{-1}\Phi_1^{1/2}-\sinh^{-1}\Phi_0^{1/2}}{\Phi_1^{1/2}\left(1+\Phi_1\right)^{-1/6}}+\frac{\myR^2}{2\myRs\myF^{1/3}}+\Oh(\myF^{-2/3}),
\end{equation}
where $\Phi_1$, $\Phi_0$, and $\myR$ are given by
\begin{align}
 \frac{1+\nu}{2}\frac{\Phi_1^{3/2}}{(1+\Phi_1)^{1/2}}&\sim  \frac{\left[\Phi^{1/2}(1+\Phi)^{1/2}-\sinh^{-1}\Phi^{1/2} \right]_{\Phi_0}^{\Phi_1}}{1-\myR^2},\\
 \frac{1+\nu}{2}\frac{\Phi_1^{3/2}}{(1+\Phi_1)^{1/2}}&\sim \frac{\Phi_0^{3/2}}{\left(\myRs^3\myF\right)^{1/2}},\\
 \frac{\myR^2}{4\left(\myRs^3\myF\right)^{1/2}}&\sim(2+\Phi_0)^{1/2}-(1+\Phi_0)^{1/2}.
\end{align}
\end{subequations}
An asymptotic expansion in $\Phi_0\to 0$ (this is equivalent to taking $\myR\to 0$) leads to the leading-order solutions \eqref{eq:cubic_cyl} for a cylindrical indenter and \eqref{eq:cubic_sph_full} for a spherical-capped indenter.   [Note that although $\Phi_0=\Phi_1=0$ is a solution of \myeqref{eq:cubic_cyl_full_app}{b,c}, this does not correspond to the limit of large indentation depth. A local expansion around $\Phi_0=\Phi_1=0$ shows that this is not the correct solution unless $\myR^2=(3\nu-1)/(\nu+1)$, as discussed by \cite{Vella2017} for the case of a point indenter ($\myR=0$); they noted that the trivial solution is only viable if $\nu=1/3$. Instead, equations \myeqref{eq:cubic_cyl_full_app}{b} and \myeqref{eq:cubic_sph_full_app}{b} must be divided through by $\Phi_1^{3/2}$.]

\section{Choice of strain energy density function}
\label{app:strain_energy}

The construction of a strain energy density function,  $W(\lambda_1, \lambda_2, \lambda_3)$, is a  difficult process,  being dependent on both the material and experiment at hand (see discussion in \eg~\citealt[Chp.~6]{Holzapfel2002}). Here, our focus is on a qualitative representation of hyperelasticity, and so  we seek  suitable strain energy density functions. In this Appendix, we construct the strain energy function used in this paper.

We require our stresses to be consistent with Hookean elasticity in the small strain limit, \ie~$T_\xi\sim\sigma_{rr}$ and $T_\phi\sim\sigma_{\theta\theta}$ in the limit $\lambda_\xi$, $\lambda_\phi\to 1$, for $T_\xi$ and $T_\phi$  defined in \eqref{eq:stress_stretch}. Taylor expanding \eqref{eq:stress_stretch} around $\lambda_\xi=\lambda_\phi=1$ (see \eg~\citealt{Horgan2004}), provides the consistency conditions
\begin{subequations}\label{eq:strain_consist}
\begin{align}
W_{I_1} +W_{I_2} =-\left(W_{I_2} +W_{I_3}\right)&= \frac{\mu}{2},\\
W_{I_1I_1} + 4W_{I_1I_2} + 2W_{I_1I_3} + 4W_{I_2I_2}  + 4W_{I_2I_3} + W_{I_3I_3}
 &= \frac{\lambda}{4}+\frac{\mu}{2},
\end{align}
\end{subequations}
for Lam\'e parameters $\mu$ and $\lambda$, and tensor invariants
\begin{subequations}\label{eq:tensor_invar}
\begin{equation}
I_1\coloneqq\lambda_1^2 +\lambda_2^2+\lambda_3^2, \quad I_2\coloneqq\lambda_1^2\lambda_2^2 +\lambda_3^2\lambda_1^2+\lambda_2^2\lambda_3^2 , \quad I_3\coloneqq\lambda_1^2\lambda_2^2\lambda_3^2, \subtag{a--c}
\end{equation}
\end{subequations}
where we define
\begin{subequations}
\begin{equation}
W_{I_i} \coloneqq \frac{\partial W}{\partial I_i}\Bigg|_{\lambda_1=\lambda_2=\lambda_3=1} \text{and} \qquad  W_{I_i I_j} \coloneqq \frac{\partial^2 W}{\partial I_i\partial I_j}\Bigg|_{\lambda_1=\lambda_2=\lambda_3=1}.\subtag{a,b}
\end{equation}
\end{subequations}
(Provided  \eqref{eq:strain_consist} holds, our membrane equations \eqref{eq:nl_odesystem} linearize to the zero bending-stiffness F\"oppl--von K\'arm\'an equations, \eqref{eq:fvk} with $B=0$,  independent of the choice of strain energy density.)

As discussed in the main text, we  concentrate on neo-Hookean and Gent solids. To formulate their strain density functions, it is convenient to split the strain energy into an isochoric and volumetric part:
\begin{equation}
W(\lambda_1,\lambda_2,\lambda_3) \equiv W_\mathrm{iso}(I_1,I_2,I_3) + W_\mathrm{vol}(J),
\end{equation}
for the tensor invariants \eqref{eq:tensor_invar} and Jacobian $J=I_3^{1/2}$. The benefit of this is that the isochoric part contains the full hyperelastic material model; that is,
\begin{subequations}\label{eq:iso_strain}
\begin{align}
\text{Neo-Hookean:} \qquad &  W_\mathrm{iso} = \frac{\mu}{2}\bigg[(I_1-3)-\log I_3\bigg],\\
\text{Gent:} \qquad &  W_\mathrm{iso} = \frac{\mu}{2}\bigg[-\frac{1}{b}\log\left[1-b(I_1-3)\right]-\log I_3\bigg],\label{eq:iso_gent}
\end{align}
\end{subequations}
for some empirical parameter $b$ based on the finite extensibility limit of the material ($1/b \equiv \max\{ I_1 -3\}$). Since the neo-Hookean model is recovered as  $b\to0$ in the Gent model, we consider only the Gent formulation henceforth. The volumetric part is a requirement of compressible materials ($J\neq 1$) and can take many constitutive forms \citep{Holzapfel2002,Horgan2004}. We take
\begin{equation}\label{eq:vol_strain}
W_\mathrm{vol} = c_1\log J +c_2(\log J)^2 +c_3 (J^2-1),
\end{equation}
for constants $c_i$; the $c_i$ are determined by inserting the volumetric strain energies \eqref{eq:vol_strain} with  Gent isometric strain \eqref{eq:iso_gent} into the consistency relations \eqref{eq:strain_consist} to give:
\begin{equation}\label{eq:vol_const}
 c_1 = -2 c_3 = c_2 +b\mu -\frac{\lambda}{2},
\end{equation}
where we are free to choose $c_2$.  Without loss of generality, we choose   $c_2=0$ (to suppress the $(\log J)^2$ term) --- this form was first proposed by \cite{Simo1992} in the context of thermoplasticity. Coupling with \eqref{eq:iso_gent},  our  strain energy density function takes the final form
\begin{equation}\label{eq:final_strain_energy}
W = \frac{\mu}{2}\bigg[-\frac{1}{b}\log\big[1-b(I_1-3)\big] +(\beta-b)(I_3-1)-(1+\beta-b)\log I_3\bigg],
\end{equation}
for $\beta \coloneqq \lambda/2\mu = \nu/(1-2\nu)$.

Although \eqref{eq:final_strain_energy} is  not necessarily the simplest choice of strain energy function, it does contain some nice properties for our work. These include:  compressibility (\ie~a variable Poisson's ratio $\nu$), which is needed to match up with the general F\"oppl--von K\'arm\'an equations;  a variable nonlinearity (through the finite-extensibility parameter $b$), giving us the ability to continuously vary our hyperelastic material model and include the common neo-Hookean hyperelasticity as a sub-case ($b=0$); the finite-extensibility property of Gent ($b>0$) could qualitatively model properties seen in complex materials such as graphene and $\mathrm{MoS}_2$ (specifically the finite length of bonds between atoms).


\end{document}